\documentclass[11pt]{article}

\usepackage{float,enumitem,amsthm,amsmath,amsfonts,amssymb,graphicx,latexsym,mathrsfs}

\usepackage{graphicx}

\usepackage{fullpage,setspace,pgfplots}

\theoremstyle{plain}
\newtheorem{theorem}{Theorem}[section]
\newtheorem{lemma}[theorem]{Lemma}
\newtheorem{proposition}[theorem]{Proposition}
\theoremstyle{definition}

\newtheorem{example}{Example}[section]
\newtheorem{assumption}{Assumption}[section]
\theoremstyle{remark}
\newtheorem*{remark}{Remark}

\newcommand{\tp}{l}
\newcommand{\TP}{L}
\newcommand{\BN}{S}
\newcommand{\SV}{\mathcal A}

\newcommand{\calN}{{\mathcal N}}
\newcommand{\calC}{{\mathcal C}}

\newcommand{\sumiN}{\sum_{i = 1}^{N}}

\newcommand{\expect}[1]{{\mathbb E}\{{ #1 }\}}
\newcommand{\indi}[1]{{\rm I}_{\{{ #1}\}}}
\newcommand{\pr}[1]{{\mathbb P}\{{ #1 }\}}

\DeclareMathOperator{\argmin}{argmin}

\begin{document}

\title{Delay Performance and Mixing Times in Random-Access Networks\thanks{This work was supported by Microsoft Research through its PhD Scholarship Programme, by the European Research Council (ERC) and by the Netherlands Organisation for Scientific Research (NWO).}}
%\author{N. Bouman\thanks{Eindhoven University of Technology, P.O. Box 513, 5600 MB Eindhoven, The Netherlands.} \and S.C. Borst\footnotemark[1] \thanks{Alcatel-Lucent Bell Labs, P.O. Box 636, Murray Hill, NJ 07974-0636, USA.} \and J.S.H. van Leeuwaarden\footnotemark[1]}
\author{
Niek Bouman, Sem C. Borst, Johan S.H. van Leeuwaarden\\
       {Eindhoven University of Technology}\\
       {Den Dolech 2, 5612 AZ Eindhoven, The Netherlands}\\
       {\{n.bouman, s.c.borst, j.s.h.v.leeuwaarden\}@tue.nl}
}

\date{}

\maketitle

\begin{abstract}
We explore the achievable delay performance in wireless random-access networks. While relatively simple and inherently distributed in nature, suitably designed queue-based random-access schemes provide the striking capability to match the optimal throughput performance of centralized scheduling mechanisms in a wide range of scenarios. The specific type of activation rules for which throughput optimality has been established, may however yield excessive queues and delays.

Motivated by that issue, we examine whether the poor delay performance is inherent to the basic operation of these schemes, or caused by the specific kind of activation rules. We derive delay lower bounds for queue-based activation rules, which offer fundamental insight in the cause of the excessive delays. For fixed activation rates we obtain lower bounds indicating that delays and mixing times can grow dramatically with the load in certain topologies as well.

%\keywords{Delay Performance \and Mixing Times
%\and Random-Access Algorithms \and Wireless Networks}
%\PACS{PACS code1 \and PACS code2 \and more}
%\subclass{MSC code1 \and MSC code2 \and more}
\end{abstract}

\section{Introduction}

Emerging wireless mesh networks typically lack any centralized access
control entity, and instead vitally rely on the individual nodes to
operate autonomously and fairly share the medium in a distributed fashion.
%This requires the nodes to schedule their individual transmissions
%and decide on the use of a shared medium based on knowledge that is
%locally available or only involves limited exchange of information.
A particularly popular mechanism for distributed medium access control
is provided by the so-called Carrier-Sense Multiple-Access (CSMA) protocol.
%various incarnations of which are implemented in IEEE 802.11 networks.
In the CSMA protocol each node attempts to access the medium after
a certain random back-off time, but nodes that sense activity of
interfering nodes freeze their back-off timer until the medium is
sensed idle.

While the CSMA protocol is fairly easy to understand at a local level,
the interaction among interfering nodes gives rise to quite intricate
behavior and complex throughput characteristics on a macroscopic scale.
In recent years relatively parsimonious models have emerged that
provide a useful tool in evaluating the throughput characteristics
of CSMA-like networks.
%These models essentially assume that the interference constraints
%can be represented by a general conflict graph, and that the
%various nodes activate asynchronously whenever none of their
%neighbors are presently active.
These models were originally developed by Boorstyn
{\em et al.}~\cite{BKMS87}, and further pursued by Wang \& Kar~\cite{WK05},
Durvy {\em et al.} \cite{DDT07,DT06} and Garetto {\em et al.}~\cite{GSK08}.
%These models in fact long pre-date the IEEE 802.11 standard and were
%already considered in the 1980's \cite{BK80,BKMS87,Kelly85,KBC87}.
%The model has strong connections with Markov random fields
%and migration processes, and can under certain assumptions be
%interpreted as a special instance of a loss network
%\cite{Kelly79,Kelly87,Kelly91,SR04,ZZ99}.
Although the representation of the CSMA back-off mechanism in the
above-mentioned models is less detailed than in the landmark work of
Bianchi~\cite{Bianchi00}, they accommodate a general interference
graph and thus cover a broad range of topologies.
Experimental results of Liew {\em et al.}~\cite{LKLW08} demonstrate
that these models, while idealized, provide throughput estimates that
match remarkably well with measurements in actual real-life networks.

Despite their asynchronous and distributed nature, CSMA-like
algorithms have been shown to offer the capability of achieving the
full capacity region and thus match the optimal throughput performance
of centralized scheduling mechanisms operating in slotted time,
see for instance Jiang \& Walrand~\cite{JW10},
Liu {\em et al.}~\cite{LYPCP08} and Tassiulas \& Ephremides~\cite{TE92}.
%More specifically, any throughput vector inside the convex hull
%associated with the independent sets in the interference graph can be
%achieved through suitable back-off rates and/or transmission lengths.
Based on this observation, various clever algorithms have been developed
for finding the back-off rates that yield a particular target throughput
vector or that optimize a certain concave throughput utility function
in scenarios with saturated buffers, see for instance
Jiang {\em et al.}~\cite{JSSW10,JW10} and Marbach \& Eryilmaz~\cite{ME08}.

In the same spirit, several powerful approaches have been devised for
adapting the transmission periods based on the queue lengths in
non-saturated scenarios, see for instance
Rajagopalan {\em et al.}~\cite{RSS09a}, Shah \& Shin~\cite{SS12}
and Shah {\em et al.}~\cite{SST11}.
Roughly speaking, the latter algorithms provide maximum-stability
guarantees under the condition that the transmission durations of the
various nodes behave as logarithmic functions of the queue lengths.

Unfortunately, however, simulation experiments demonstrate that such
activation rules can induce excessive queues and delays,
which has sparked a strong interest in developing approaches for
improving the delay performance, see for instance Ghaderi
\& Srikant~\cite{GS10}, Lotfinezhad \& Marbach~\cite{LM10},
Ni {\em et al.}~\cite{NTS10} and Shah \& Shin~\cite{SS10}.
In particular, it has been shown that more aggressive schemes,
where the transmission durations grow faster as function of the
queue lengths, can reduce the delays, see for instance Bouman
{\em et al.}~\cite{BBBvL12}.

%Motivated by this issue, Ghaderi \& Srikant~\cite{GS10} recently
%showed that it is in fact sufficient for the \emph{logarithms} of the
%activity factors to behave as logarithmic functions of the queues,
%divided by an arbitrarily slowly increasing, unbounded function.
%These results indicate that the maximum-stability guarantees are
%preserved for activity functions that are essentially linear for all
%practical values of the backlogs, although asymptotically the activity
%rate must grow slower than any positive power of the backlog.
%A careful inspection reveals that the proof arguments leave little
%room to weaken the stated growth condition.
%Since the growth condition is only a sufficient one, however, it is
%not clear to what extent it is actually a strict requirement for
%maximum stability to be maintained.

In order to gain insight in the root cause for the poor delay performance,
we establish in the present paper lower bounds for the average
steady-state delay.
%The bounds indicate that the delay can dramatically grow with the
%load in certain topologies, even with low node degrees.
To the best of our knowledge, the derivation of lower bounds for the
average steady-state delay in random-access networks has received
hardly any attention so far.
An interesting paper by Shah {\em et al.}~\cite{STT11} showed that
low-complexity schemes cannot be expected to achieve low delay in
arbitrary topologies (unless P equals NP), since that would imply that
certain NP-hard problems could be solved efficiently.
However, the notion of delay in~\cite{STT11} is a transient one,
and it is not exactly clear what the implications are for the
average steady-state delay in specific networks, if any.

Jiang {\em et al.}~\cite{JLNSW11,JW12} derived {\em upper\/} bounds
for the average steady-state delay based on mixing time results for
Glauber dynamics, where the mixing time represents the amount of time
required for the process to come close to its equilibrium distribution.
The bounds show that for sufficiently low load the delay only grows
polynomially with the number of nodes in bounded-degree interference
graphs.
Subramanian \& Alanyali~\cite{SA11} presented similar
upper bounds for bounded-degree interference graphs with low load
based on analysis of neighbor sets and stochastic coupling arguments.
While some of the conceptual notions in the present paper are
similar (cliques, mixing times), we focus on {\em lower\/} rather
than upper bounds, and exploit quite different techniques.

The lower bounds that we derive for queue-based activation schemes
provide fundamental insight why the kind of rules that guarantee
maximum stability yield excessive delays.
We further obtain lower bounds for the delay and mixing time in case
fixed back-off rates are used.
In both cases, the bounds bring to light that the delay and mixing
time can grow dramatically with the load of the system.
Specifically, we establish that the expected delay grows as
$F(1 / (1 - \rho))$ as $\rho \uparrow 1$, where $\rho$ is the load
and $F(\cdot)$ is a superlinear function, implying that the growth
rate may be polynomially or even exponentially faster than is
typically the case in queueing systems at high load.
The specific form and growth rate of the function $F(\cdot)$ depends
on the activation rule as well as the topology of the network,
as we will show for several scenarios of interest.
Various partial versions of the results presented here appeared in
Bouman {\em et al.} \cite{BBL11a,BBLP11}.

The remainder of the paper is organized as follows.
In Section~\ref{mode} we present a detailed model description,
followed by some preliminary results in Section~\ref{prel}.
In Section~\ref{backlogbased} we derive delay lower bounds for
queue-based activation schemes.
We establish generic lower bounds for the delay and mixing time in
case of fixed back-off rates in Section~\ref{debo}.
In Sections~\ref{sscompl} and~\ref{exte} we apply these generic bounds to
a canonical class of partite interference graphs, which includes
several specific cases of interest such as grid topologies.
Simulation experiments are conducted in Section~\ref{simu} to support
the analytical results.
In Section~\ref{concl}, we make some concluding remarks and identify
topics for further research.

\section{Model description}
\label{mode}

\paragraph{Network, interference graph, and traffic model.}

We consider a network of several nodes sharing a wireless medium
according to a random-access mechanism.
The network is represented by an undirected graph $G = (V , E)$
where the set of vertices $V = \{1,\ldots, N\}$ correspond to the
various nodes and the set of edges $E \subseteq V \times V$ indicate
which pairs of nodes interfere.
Nodes that are neighbors in the interference graph are prevented
from simultaneous activity, and thus the independent sets of~$G$
correspond to the feasible joint activity states of the network.
A node is said to be blocked whenever the node itself or any of
its neighbors is active, and unblocked otherwise.
Define $\Omega \subseteq \{0, 1\}^N$ as the set of all feasible joint
activity states of the network.
%the incidence vectors of the independent sets of the interference graph.

Packets arrive at node~$i$ as a Poisson process of rate~$\lambda_i$.
The packet transmission times at node~$i$ are independent
and exponentially distributed with mean $1 / \mu_i$.
Denote by $\rho_i = \lambda_i/\mu_i$ the traffic intensity of node~$i$.

Let $U(t) \in \Omega$ represent the joint activity state of the network
at time~$t$, with $U_i(t)$ indicating whether node~$i$ is active at
time~$t$ or not.
Denote by $\TP_i(t)$ the number of packets at node~$i$ at time~$t$ (including any packet that may be in the
process of being transmitted).

\paragraph{Random-access mechanism.}

The nodes share the me\-dium according to a random-access mechanism.
When a node ends an activity period (consisting of possibly several
back-to-back packet transmissions), it starts a back-off period.
The back-off times of node~$i$ are independent and exponentially
distributed with mean $1 / \nu_i$.
The back-off period of a node is suspended whenever it becomes
blocked by activity of any of its neighbors, and is only resumed once
the node becomes unblocked again.
Thus the back-off period of a node can only end when none of its
neighbors are active.
Now suppose a back-off period of node~$i$ ends at time~$t$.
Then the node starts a transmission with probability $\phi_i(\TP_i(t))$,
and begins a next back-off period otherwise.
When a transmission of node~$i$ ends at time~$t$, it releases the
medium and begins a back-off period with probability $\psi_i(\TP_i(t^+))$,
or starts the next transmission otherwise.
We allow for $\phi_i(0) > 0$ and $\psi_i(0)<1$, so a node may be
active even when its buffer is empty, and transmit dummy packets.
A dummy transmission is terminated when a new packet arrives and the
transmission of this packet is started immediately.
Equivalently, node~$i$ may be thought of as activating at
an exponential rate $f_i(\TP_i(t))= \nu_i \phi_i(\TP_i(t))$,
whenever it is unblocked at time~$t$, and de-activating at rate
$g_i(\TP_i(t)) = \mu_i \psi_i(\TP_i(t)-1)$, whenever it is active
at time~$t$.
For conciseness, the functions $f_i(\cdot)$ and $g_i(\cdot)$ will be
referred to as activation and de-activation functions, respectively,
and we define $h_i(\cdot)=f_i(\cdot)/g_i(\cdot)$ as the nominal
activation function.

\paragraph{Network dynamics.}

Under the above-described queue-based schemes, the process
$\{(U(t), \TP(t))\}_{t \geq 0}$ evolves as a Markov process with state
space $\Omega \times {\mathbb N}_0^N$.
Transitions (due to arrivals) from a state $(U, \TP)$ to $(U, \TP + e_i)$
occur at rate~$\lambda_i$, transitions (due to activations) from
a state $(U, \TP)$ with $U_i = 0$ and $U_j = 0$ for all neighbors~$j$ of
node~$i$ to $(U + e_i, \TP)$ occur at rate
$f_i(\TP_i)$, transitions (due to transmission completions
followed back-to-back by a subsequent transmission) from a state
$(U, \TP)$ with $U_i = 1$ to $(U, \TP - e_i \indi{\TP_i > 0})$ occur at
rate $\mu_i - g_i(\TP_i)$, transitions (due to transmission
completions followed by a back-off period) from a state $(U, \TP)$
with $U_i = 1$ to $(U - e_i, \TP - e_i \indi{\TP_i > 0})$ occur at rate $g_i(\TP_i)$.

\paragraph{Product-form distribution.}

We now proceed with some additional notation and preliminary results.
For any $u \in \Omega$, define
$\pi(u) = \lim_{t \to \infty} \pr{U(t) = u}$ as the steady-state
probability that the activity process resides in state~$u$.
Further define $\theta_i = \sum_{u \in \Omega} \pi(u) u_i$ as the
steady-state fraction of time that node~$i$ is active.
Note that for fixed activation and de-activation rates,
i.e., $\phi_i(\cdot) \equiv \phi_i$ and $\psi_i(\cdot) \equiv \psi$,
the activity process $\{U(t)\}$ does not depend on the process
$\{\TP(t)\}$, and in fact constitutes a reversible Markov process
with product-form stationary distribution~\cite{BKMS87}
\begin{equation}
\label{statdistU}
\pi(u) = Z^{- 1} \prod\limits_{i = 1}^{N} \sigma_i^{u_i},
\hspace*{.4in} u \in \Omega,
\end{equation}
and normalization constant
\[
Z = \sum_{u \in \Omega} \prod\limits_{i = 1}^{N} \sigma_i^{u_i},
\]
with $\sigma_i = \nu_i \phi_i / (\mu_i \psi_i)$ representing a nominal
activity factor.

\paragraph{Stability.}

In general it is difficult to establish under what conditions the
system is stable, i.e., when the process $\{(U(t), \TP(t))\}_{t \geq 0}$
is positive recurrent.
Denoting by $\mbox{conv}(\cdot)$ the convex hull operator
and by $\mbox{int(conv}(\cdot))$ its interior, it is easily seen that
$(\rho_1, \dots, \rho_N) \in \mbox{int(conv}(\Omega))$ is a necessary
condition for stability.

In \cite{GS10,RSS09a,SS12} it is shown that this condition is in fact
also sufficient for activation and de-activation functions
$f_i(\tp) = r_i(\tp) / (1 + r_i(\tp))$ and $g_i(\tp) = 1 / (1 + r_i(\tp))$
with suitably chosen $r_i(\cdot)$, e.g., $r_i(\tp) = \log(\tp + 1)$.
For more aggressive queue-based activation functions, \cite{GBW12}
shows that the necessary condition is not always sufficient though.
%Finding the exact characterization of the stability region, depending on the structure of the network and the activity functions, is still an open problem. For $f_i(\tp) \equiv \nu_i$, $\tp \geq 1$, $f_i(0) = 0$, $g_i(\tp) \equiv 1$,
%\cite{VBLP10} derives necessary and sufficient stability conditions
%for full interference graphs, and shows that an exact characterization
%of the stability region is difficult for all other topologies in this case.

In the case of fixed activation and de-activation rates, a simple
necessary and sufficient condition for stability is $\rho_i < \theta_i$,
for all $i = 1, \dots, N$.
Furthermore, there exists a unique vector $(\sigma_1,\dots,\sigma_N)$
that yields $(\theta_1,\dots,\theta_N) \in \mbox{int(conv}(\Omega))$
\cite{JW10,VJLB11}.
Hence, for any traffic intensity vector obeying the necessary stability
condition, $(\rho_1, \dots, \rho_N) \in \mbox{int(conv}(\Omega))$,
there exists a vector $(\sigma_1,\dots,\sigma_N)$ such that
$(\rho_1, \dots, \rho_N) < (\theta_1,\dots,\theta_N) \in
\mbox{int(conv}(\Omega))$, though determining the right vector
$(\sigma_1,\dots,\sigma_N)$ is non-trivial in general.
%One could use the global invertibility~\cite{VJLB11} of the throughput function, but this is computationally heavy and requires knowledge of the topology and arrival rates. A distributed algorithm for finding $(\sigma_1,\dots,\sigma_N)$ was proposed in~\cite{JW10} and shown to converge in~\cite{JSSW10}.
%While the proof of convergence in~\cite{JW10} is based on a time-scale separation assumption, the authors in~\cite{JSSW10} show without this assumption that the proposed algorithm converges. Unfortunately this convergence is slow, which can potentially have a huge impact on the (short-term) delay.

\section{Preliminary results}
\label{prel}

In this section we state some preliminary results in preparation for
the derivation of delay lower bounds in the next sections.
Throughout we assume that the system under consideration is stable,
because otherwise such lower bounds are not particularly meaningful.
More specifically, we derive lower bounds for the expected aggregate
stationary queue length in subsets of nodes $\SV \subseteq V$.
Note that, using Little's law, this also provides a lower bound for
the expected aggregate stationary delay.
That is, $\sum_{i \in \SV} \expect{\TP_i} \geq \alpha$ implies that
$\sum_{i \in \SV} \lambda_i \expect{W_i} \geq \alpha$, with $W_i$
a random variable representing the delay (waiting time plus service time)
of an arbitrary packet at node~$i$.
%Further note that the total number of packets in an unstable system is infinitely large, so that lower bounds are not relevant for that case.

The notion of a {\em clique\/} will play a pivotal role in the
derivation of the lower bounds.
A clique is a subset $\calC \subseteq V$ of vertices in the interference
graph~$G$ such that the subgraph induced by $\calC$ is complete.
Note that in a clique at most one node can be active at a time.
The aggregate load in a clique should therefore be less than one
if the system is to be stable.
For compactness, we use the notation
$\lambda_{\calC} = \sum_{j \in \calC} \lambda_j$
and $\rho_{\calC} = \sum_{j \in \calC} \rho_j$.
We say that a clique $\calC$ is in heavy traffic when $\rho_{\calC}$
is close to one.
Further we denote by $\TP_{i, \calC}$ the number of packets at node~$i$ at an arbitrary epoch during a non-serving interval for the clique~$\calC$, i.e., a time interval during which none of the nodes in~$\calC$ is transmitting a packet.
%we are especially interested in the delay behavior of nodes that belong to a clique that is in heavy traffic.

Observe that the total number of packets in the clique~$\calC$
is bounded from below by that in a single node carrying the aggregate
traffic, yielding the simple lower bound
\begin{equation}
\label{LBFC}
\sum_{i \in \calC} \expect{\TP_i} \geq
\frac{\lambda_{\calC} \sum_{i \in \calC} \lambda_i / \mu_i^2}{1 - \rho_{\calC}} + \rho_{\calC}.
\end{equation}
Thus, the total expected number of packets in any system grows
at least linearly in $1/(1-\rho)$ as $\rho$ increases to one,
with $\rho=\max_{\calC} \rho_{\calC}$ the maximum traffic intensity
in any clique. \\

The lower bound in~\eqref{LBFC} is only based on sheer load considerations
and does not account for the effect of the back-off mechanism.
In the next sections we will derive lower bounds for queue-based
strategies as well as fixed-rate strategies that do capture the effect
of the back-off mechanism, and turn out to be considerably tighter
and exhibit superlinear growth in $1 / (1 - \rho_{\calC})$.

The derivation of the lower bounds starts from the observation that
stability of the system requires the non-serving intervals for a clique
in heavy traffic to be short or happen infrequently.
That is, in each clique, most of the time, one of the nodes should be
active, since otherwise the average rate of arriving packets would
exceed the average rate of departing packets.
For this to be the case, the activity factors should be big at high load.
The next lemma quantifies this statement.

\begin{lemma}
\label{lbactfac}
Assume that the system is stable. Then, for any clique $\calC \subseteq V$
containing node~$i$,
\begin{equation}
\label{relbacklogb}
\frac{\expect{f_i(\TP_{i, \calC})}}{\expect{g_i(\TP_i)}} \geq
\frac{\rho_i}{1-\rho_{\calC}} .
\end{equation}
%${\rm (ii)}$ If $\sigma_j(\cdot)=\nu_i\phi_i(\cdot)/(\mu_i\psi_i(\cdot)) \equiv \sigma_j$ for all $j \in V$,
%\begin{equation}
%\sigma_i > \frac{\rho_i}{1 - \rho_{\calC}}.
%\label{ineq5}
%\end{equation}
\end{lemma}

\textbf{Proof}
Observing that the mean number of activations at node~$i$ equals
the mean number of de-activations at node~$i$ per unit of time,
%provided node~$i$ is stable,
we obtain
\begin{equation}\label{actisdeact}
\expect{f_i(\TP_i) \indi{U_j = 0 \mbox{\scriptsize ~for all } j \in \calN_i^+}} =
\rho_i \expect{g_i(\TP^d_i)},
\end{equation}
where $\calN_i^+$ denotes the set of neighbors of node~$i$ in the
graph~$G$, along with~$i$ itself, and $\TP^d_i$ denotes the number of
packets waiting for transmission at node~$i$ at a departure epoch.
Note that $\TP^d_i$ is, in distribution, equal to $\TP_i$.

Further,
\begin{align}
 \expect{f_i(\TP_i) \indi{U_j = 0 \mbox{\scriptsize ~for all } j \in \calN_i^+}} %\nonumber \\
&
\leq
\expect{f_i(\TP_i) \indi{U_j = 0 \mbox{\scriptsize ~for all } j \in \calC}} \nonumber \\
&
 =
\expect{f_i(\TP_{i, \calC})} \pr{U_j = 0 \mbox{\scriptsize ~for all } j \in \calC}. \label{boundact}
\end{align}
%\begin{align}
% &\expect{f_i(\TP_i) \indi{U_j = 0 \mbox{\scriptsize ~for all } j \in \calN_i^+}} \leq
%\expect{f_i(\TP_i) \indi{U_j = 0 \mbox{\scriptsize ~for all } j \in \calC}} \nonumber \\
% & \hspace{2cm} =
%\expect{f_i(\TP_{i, \calC})} \pr{U_j = 0 \mbox{ for all } j \in \calC}. \label{boundact}
%\end{align}
Since the events $\{U_j = 1\}$ are mutually exclusive for all
$j \in \calC$, it follows that
\begin{align}
\pr{U_j = 0 \mbox{ for all } j \in \calC} &= 1 - \pr{U_j = 1 \mbox{ for some } j \in \calC} \nonumber \\
&=1 - \sum_{j \in \calC} \pr{U_j = 1} = 1 - \sum_{j \in \calC} \theta_j. \label{actinclique}
\end{align}
Thus, we find~\eqref{relbacklogb} using~\eqref{actisdeact},
\eqref{boundact}, and the fact that $\rho_i \leq \theta_i$ for all
$i \in V$ is a necessary condition for stability of the system.
%If $\sigma_j(\cdot) \equiv \sigma_j$ we know from the product-form stationary distribution~\eqref{statdistU} that
%\[
%\theta_i=\sigma_i \pr{U_j = 0 \mbox{ for all } j \in \calN_i^+}.
%\]
%Furthermore,
%\[
%\pr{U_j = 0 \mbox{ for all } j \in \calN_i^+} \leq \pr{U_j = 0 \mbox{ for all } j \in \calC},
%\]
%and we hence get
%\begin{equation}
%\label{thetab}
%\theta_i \leq \sigma_i [1 - \sum_{j \in \calC} \theta_j]
%\end{equation}
%from~\eqref{actinclique}. Thus, as $\rho_i < \theta_i$ for all $i \in V$ is a necessary condition for stability of the system, we find~\eqref{ineq5}.
\qed

In particular, for fixed-rate strategies it follows that stability
entails
\begin{equation}
\sigma_i \geq \frac{\rho_i}{1 - \rho_{\calC}},
\label{ineq5}
\end{equation}
which in fact could also have been established using the product-form
distribution~\eqref{statdistU}. \\

Lemma~\ref{lbactfac} shows that the activity factors in each clique
should be big at high load.
In the next sections we will demonstrate that this also causes the
delay and mixing time to grow dramatically in heavy traffic.

\paragraph{Queue-based strategies.}

For queue-based strategies we examine in Section~\ref{backlogbased}
activation functions that are such that a node becomes increasingly
more aggressive when the total number of packets at that node increases.
%In that case we see that the delay performance is optimal when $\TP_i$
%would be constant, or concentrated around its mean,
%see Theorem~\ref{backlogb}.
For that natural class of activations functions, we exploit the result of Lemma~\ref{lbactfac} to find a lower bound of the form $h^{-1}(1/(|\calC|(1-\rho_{\calC})))$ for the aggregate number of packets in the clique~$\calC$, where $h^{-1}(\cdot)$ is the inverse function of $h(\cdot)$.

A prominent example is $f(\tp) = r(\tp) / (1 + r(\tp))$
and $g(\tp) = 1 / (1 + r(\tp))$ with $r(\tp) = \log(\tp + 1)$,
so that $h^{-1}(\tp) = \exp(\tp) - 1$, the class of backlog-based
strategies for which maximum stability is guaranteed as mentioned earlier.
In this case we find that the queue length scales at least exponentially
in $1/(1-\rho_{\calC})$.
%To illustrate the performance ramifications, note that for $\rho=0.9$, this exponential growth factor is already a factor 2000 larger than the normal growth factor at that load, while at larger loads it is even worse!

\paragraph{Fixed-rate strategies.}

In the case of fixed-rate strategies the delay lower bounds revolve
around two simple observations:
(i) high activation rates cause long mixing times,
in particular slow transitions between dominant activity states;
(ii) slow transitions between dominant states imply long starvation periods for some nodes,
and hence huge queue lengths and delays.
In Section~\ref{debo} we formalize (ii), and establish lower bounds for
the expected aggregate weighted queue length and delay in terms
of the expected return times of the process $\{U(t)\}$.

In order to lower bound these return times, we will build
in Sections~\ref{sscompl} and~\ref{exte} on insight~(i) for a canonical class of partite
interference graphs.
That is, we examine topologies where the nodes belong to one of
$K$~different components such that nodes in the same component do not
interfere with each other and every node belongs to a clique of size~$K$
(of which the other $K-1$ nodes necessarily belong to $K-1$
different components).
This class of $K$-partite interference graphs covers a wide range
of network topologies with nearest-neighbor interference,
e.g., linear topologies, ring networks with an even number of nodes,
two-dimensional grid networks, tori (two-dimensional grid networks
with a wrap-around boundary), and {\em complete\/} $K$-partite graphs,
where all nodes are connected except those that belong to the same
component, with star topologies as a prime example.

\section{Queue-based strategies}
\label{backlogbased}

%Over the last few years, significant interest has emerged in
%queue-based activation and de-activation strategies.
%As mentioned earlier, such strategies, when suitably designed,
%can achieve throughput optimality without requiring explicit knowledge of the arrival rates.
%However, simulations suggest that the type of activation rules for which throughput optimality
%has been established may yield excessive delay and queues.

In this section we derive delay lower bounds for queue-based strategies
that use a concave activation function or a convex de-activation function.
%For compactness, we use the notation $\xi_{\calC} = \sum_{j \in \calC} \xi_j$ and $\sum_{j \in \calC} \rho_j\xi_j = \sum_{j \in \calC} \rho_j\xi_j$.

\begin{theorem}
\label{backlogb}
Assume $\lambda_i$, $\nu_i$, $\mu_i$, $f_i(\cdot)$ and $g_i(\cdot)$, $i=1,\dots,N$, are such that the system is stable. Then, for any clique $\calC \subseteq V$,

${\rm (i)}$ If $f_i(\cdot) \equiv f(\cdot)$ for $i \in \calC$ is an increasing concave function and $g_i(\cdot)\geq \xi_i >0$ for $i \in \calC$, then
\begin{equation}
\label{backlogbc1}
\sum_{i \in \calC} \expect{\TP_i} \geq
\frac{\lambda_{\calC} \sum_{i \in \calC} \lambda_i / \mu_i^2}{1 - \rho_{\calC}} +
|\calC| f^{- 1}\Big(\frac{1}{|\calC|}
\frac{\sum_{i \in \calC} \rho_i\xi_i}{1 - \rho_{\calC}}\Big) + \rho_{\calC}.
\end{equation}

${\rm (ii)}$ If $f_i(\cdot)\leq \xi_i$ for $i \in \calC$ and $g_i(\cdot) \equiv g(\cdot)$ for $i \in \calC$ is a decreasing convex function, then
\begin{equation}
\label{backlogbc2}
\sum_{i \in \calC} \rho_i \expect{\TP_{i}} \geq \rho_{\calC} g^{-1} \Big(\frac{(1-\rho_{\calC})\sum_{i \in \calC} \xi_i}{\rho_{\calC}}\Big).
\end{equation}

${\rm (iii)}$ If $f_i(\cdot) \equiv f(\cdot)$ for $i \in \calC$ is an increasing
concave function and $g_i(\cdot) \equiv g(\cdot)$ for $i \in \calC$ is a decreasing convex
function, then
\begin{equation}
\label{backlogbc3}
\sum_{i \in \calC} \expect{\TP_i} \geq
h^{- 1}\Big(\frac{1}{|\calC|}
\frac{\rho_{\calC}}{1 - \rho_{\calC}}\Big).
\end{equation}
\end{theorem}

\textbf{Proof}
The Fuhrmann-Cooper decomposition property~\cite{FC85}
(applied to the total number of packets in the clique~$\calC$) implies
\begin{equation}
\label{fconcaveFC}
\sum_{i \in \calC} \expect{\TP_i} =
\frac{\lambda_{\calC} \sum_{i \in \calC} \lambda_i / \mu_i^2}{1 - \rho_{\calC}} +
\sum_{i \in \calC} \expect{\TP_{i, \calC}} + \rho_{\calC}.
\end{equation}
This corroborates (\ref{LBFC}) since the second term
in~(\ref{fconcaveFC}) is non-negative, but in case~${\rm (i)}$ that term might in fact be dominant as we now proceed to show.
From~\eqref{relbacklogb} we know that, in case~${\rm (i)}$,
\[
(1 - \rho_{\calC}) \sum_{i \in \calC} \expect{f_i(\TP_{i, \calC})} \geq
\sum_{i \in \calC} \rho_i\xi_i.
\]
Since $f(\cdot)$ is concave, it follows from Jensen's inequality that
\begin{equation}
\label{fconcaveJen}
\sum_{i \in \calC} \expect{f(\TP_{i, \calC})} \leq |\calC|
f\Big(\frac{1}{|\calC|} \sum_{i \in \calC} \expect{\TP_{i, \calC}}\Big).
\end{equation}
Because $f(\cdot)$ is increasing we thus get
\[
\sum_{i \in \calC} \expect{\TP_{i, \calC}} \geq
|\calC| f^{- 1}\Big(\frac{1}{|\calC|}
\frac{\sum_{i \in \calC} \rho_i\xi_i}{1 - \rho_{\calC}}\Big),
\]
which completes the proof for case~${\rm (i)}$.

The proof for case~${\rm (ii)}$ proceeds along similar lines.
From~\eqref{relbacklogb} we obtain
\[
\sum_{i \in \calC} \rho_i \expect{g(\TP_i)} \leq (1-\rho_{\calC})\sum_{i \in \calC} \xi_i.
\]
Since $g(\cdot)$ is convex, it follows from Jensen's inequality that
\begin{equation}
\label{gconvexJen}
\sum_{i \in \calC} \rho_i \expect{g(\TP_i)} \geq \rho_{\calC} g\Big(\frac{1}{\rho_{\calC}} \sum_{i \in \calC} \rho_i \expect{\TP_i}\Big).
\end{equation}
Since $g(\cdot)$ is decreasing we thus get
\[
\sum_{i \in \calC} \rho_i \expect{\TP_i} \geq \rho_{\calC} g^{-1} \Big(\frac{(1-\rho_{\calC})\sum_{i \in \calC} \xi_i}{\rho_{\calC}}\Big),
\]
yielding~(\ref{backlogbc2}).

To prove case~${\rm (iii)}$, note that
combining~\eqref{fconcaveFC} and \eqref{fconcaveJen} gives
\[
\sum_{i \in \calC} \expect{f(\TP_{i, \calC})} \leq |\calC|
f\Big(\frac{1}{|\calC|} \Big( \sum_{i \in \calC} \expect{\TP_i} -
\frac{\lambda_{\calC} \sum_{i \in \calC} \lambda_i / \mu_i^2}{1 - \rho_{\calC}} - \rho_{\calC} \Big)  \Big)
\]
and hence, because $f(\cdot)$ is increasing,
\[
\sum_{i \in \calC} \expect{f(\TP_{i, \calC})} \leq |\calC|
f\Big(\sum_{i \in \calC} \expect{\TP_i}\Big).
\]
Further, since $\rho_i \leq \rho_{\calC}$ for $i\in \calC$ and because
$g(\cdot)$ is decreasing we obtain from~\eqref{gconvexJen} that
\[
\sum_{i \in \calC} \rho_i \expect{g(\TP_i)} \geq \rho_{\calC} g\Big(\sum_{i \in \calC} \expect{\TP_i}\Big).
\]
From~\eqref{relbacklogb} we then find
\[
|\calC| f\Big(\sum_{i \in \calC} \expect{\TP_i}\Big) (1-\rho_{\calC}) \geq \rho_{\calC} g\Big(\sum_{i \in \calC} \expect{\TP_i}\Big),
\]
or
\[
h\Big(\sum_{i \in \calC} \expect{\TP_i}\Big) \geq \frac{1}{|\calC|}\frac{\rho_{\calC}}{1-\rho_{\calC}}.
\]
Thus as $h(\cdot)=f(\cdot)/g(\cdot)$ is increasing because $f(\cdot)$ is increasing and $g(\cdot)$ is decreasing, we get~\eqref{backlogbc3}.
\qed

The three cases covered in Theorem~\ref{backlogb} all reveal  the same
effect, namely that the mean number of packets in a clique is at least
of the order of $h^{-1}(1/(|\calC|(1-\rho_{\calC})))$, where $h^{-1}(\cdot)$ is the
inverse function of $h(\cdot)$.
In case~${\rm (ii)}$ this effect is observed because the argument
of $g^{-1}(\cdot)$ is reciprocal.
Further, noting that $f(\tp) = \log(\tp + 1) / (1 + \log(\tp + 1))$ is
an increasing concave function and $g(\tp) = 1 / (1 + \log(\tp + 1))$
is a decreasing convex function, we have
\[
\sum_{i \in \calC} \expect{\TP_i} \geq {\rm Exp} \Big(\frac{1}{|\calC|}
\frac{\rho_{\calC}}{1 - \rho_{\calC}}\Big) - 1
\]
for the class of functions for which maximum stability is guaranteed.

The results of Theorem~\ref{backlogb} suggest that in order to improve
the delay performance one should use more aggressive access schemes.
In fact, if $h(\cdot)$ is a superlinear function, i.e., if $h(\cdot)$
grows faster than linear, we find a lower bound that is loose in heavy
traffic and~\eqref{LBFC} provides a better lower bound in that case.
Remember however that maximum stability is not guaranteed in case
a superlinear function $h(\cdot)$ is used, hence the delay performance
might actually deteriorate, and even instability could occur as shown
in~\cite{GBW12}.

\section{Fixed-rate strategies}
\label{debo}

In the previous section we derived delay bounds for queue-based
activation rules and we saw that the type of activation rules for
which throughput optimality has been established yield excessive
delays and queues.
We now proceed to construct lower bounds for the expected aggregate
weighted queue length and delay in the case of fixed activation
and de-activation rates, i.e., we take $\phi_i(\cdot) \equiv \phi_i$
and $\psi_i(\cdot) \equiv \psi_i$.

We first introduce some useful notation.
Define $Q(\BN)$ as the transition rate out of the subset~$\BN \subseteq \Omega$, i.e.,
\[
Q(\BN) =
\sum_{u \in \BN} \sum_{u' \in \Omega \setminus \BN} \pi(u) q(u, u') =
\sum_{u \in \Omega \setminus \BN} \sum_{u' \in \BN} \pi(u) q(u, u'),
\]
with $q(u, u')$ denoting the transition rate from state~$u$ to
state~$u'$ of the component $\{U(t)\}$ of the Markov process as
specified in Section~\ref{mode}, i.e., $q(u, u + e_i) = \nu_i \phi_i$
and $q(u + e_i, u) = \mu_i \psi_i$, $u, u + e_i \in \Omega$.
With minor abuse of notation, denote by
$\pi(\BN) = \sum_{u \in \BN} \pi(u)$ the fraction of time that the
system resides in one of the activity states in the subset~$\BN$.
The bottleneck ratio of the subset~$\BN$ is defined as
\[
\Phi(\BN) = \frac{Q(\BN)}{\pi(\BN)}.
\]
Further define for arbitrary weights $w \in {\mathbb R}_+^N$ and for any $\SV \subseteq V$, $\BN \subseteq \Omega$,
\[
Y(w, \SV, \BN) = \max_{u \in \BN} \sum_{i \in \SV} w_i \mu_i u_i,
\]
and denote
\[
D(w, \SV, \BN) = \sum_{i \in \SV} w_i \lambda_i - Y(w, \SV, \BN).
\]
The coefficient $Y(w, \SV, \BN)$ represents the maximum aggregate
weighted service rate of the nodes in~$\SV$ when the system resides
in one of the activity states in the subset~$\BN$.
Noting that $\sum_{i \in \SV} w_i \lambda_i$ is the weighted
arrival rate of the nodes in~$\SV$, the coefficient $D(w, \SV, \BN)$
may thus be interpreted as the minimum drift in the aggregate
weighted queue length of the nodes in~$\SV$ when the system resides
in one of the activity states in the subset~$\BN$.

\begin{proposition}
\label{weighted}
For any $w \in {\mathbb R}_+^N$, $\SV \subseteq V$,
\begin{equation}
\label{propweigh}
\sum_{i \in \SV} w_i \expect{\TP_i} \geq
\frac{1}{2} \max_{\BN \subseteq \Omega} D(w, \SV, \BN) \pi(\BN) \frac{1}{\Phi(\BN)}.
\end{equation}

\end{proposition}

\textbf{Proof}
Denote by $T_{\BN}$ a random variable representing the equilibrium
return time to the subset of activity states $\Omega \setminus \BN$ and denote by $T_{\BN}^e$ a random variable representing the elapsed equilibrium lifetime of $T_{\BN}$, i.e.,
\begin{equation}
\label{prop1eq1}
\pr{T_{\BN}^e < t} =
\frac{1}{\expect{T_{\BN}}} \int_{s = 0}^{t} \pr{T_{\BN} > s} {\rm d}s.
\end{equation}
Now observe that when the system resides in one of the activity
states in~$\BN$, which is the case with probability $\pi(\BN)$,
the aggregate weighted queue length of the nodes in~$\SV$ have
experienced a drift no less than $D(w, \SV, \BN)$ for an expected
amount of time $\expect{T_{\BN}^e}$.
This observation indicates that the expected aggregate weighted
queue length of the nodes in~$\SV$ is bounded from below by
$\pi(\BN) D(w, \SV, \BN) \expect{T_{\BN}^e}$ for any choice of~$\BN$ and hence
\begin{equation}
\label{prop1e2}
\sum_{i \in \SV} w_i \expect{\TP_i} \geq
\max_{\BN \subseteq \Omega} D(w, \SV, \BN) \pi(\BN) \expect{T_{\BN}^e}.
\end{equation}
Using~\eqref{prop1eq1} we obtain
\begin{align*}
\expect{T_{\BN}^e} &= \frac{1}{\expect{T_{\BN}}} \int_{t=0}^{\infty} \int_{s = t}^{\infty} \pr{T_{\BN} > s} {\rm d}s {\rm d}t %\\
%&
=
\frac{\expect{T_{\BN}^2}}{2 \expect{T_{\BN}}} \geq
\frac{1}{2} \expect{T_{\BN}}.
\end{align*}
Finally, because $Q(\BN)$ is the expected number of times the process enters $\BN$ per unit of time and $\expect{T_{\BN}}$ is the expected amount of time the process stays in $\BN$ after entering, the expected fraction of the time the process resides in $\BN$, $\pi(\BN)$, is given by $\pi(\BN)=Q(\BN) \expect{T_{\BN}}$. Thus, $\expect{T_{\BN}} = \frac{1}{\Phi(\BN)}$, and~\eqref{propweigh} follows.
\qed

The question arises how to choose~$\BN$ such that the maximum and thus
the tightest possible lower bound in~\eqref{propweigh} is obtained.
Evidently, the more $\BN$ includes states with some of the nodes
in~$\SV$ active, the larger the potential aggregate weighted service
rate of the nodes in~$\SV$, i.e., the larger $Y(w, \SV, \BN)$,
and the smaller $D(w, \SV, \BN)$.
In other words, we need to ensure that $\BN$ excludes some of the
states with nodes in~$\SV$ active.
Indeed, if $\BN$ includes all states with maximal subsets of the
nodes in~$\SV$ active, then
$Y(w, \SV, \BN) = \max_{u \in \Omega} \sumiN \hat{w}_i \mu_i u_i$,
with $\hat{w}_i = w_i$ if $i \in \SV$ and $\hat{w}_i = 0$ otherwise.
The fact that $(\rho_1, \dots, \rho_N) \in \mbox{int(conv}(\Omega))$
then implies that $Y(w, \SV, \BN) \geq \sumiN \hat{w}_i \mu_i \rho_i =
\sumiN \hat{w}_i \lambda_i = \sum_{i \in \SV} w_i \lambda_i$,
so that $D(w, \SV, \BN) \leq 0$, yielding an irrelevant lower bound.
However, observe that the expected equilibrium return time to $\Omega \setminus \BN$, denoted $\expect{T_\BN}$,
may be small when $\BN$ includes very few states.
Hence, to obtain the sharpest possible lower bound, it may not
necessarily be optimal to exclude all the states with nodes
in~$\SV$ active from~$\BN$.
For high values of~$\nu$, which are necessary for stability at high
load as Lemma~\ref{lbactfac} showed, the above argument suggests
that we should choose $\BN$ so that it contains a state with many active nodes, while the boundary
of~$\BN$ only contains states with few active nodes.

Define
\[
\partial \BN = \{u \in \BN: \sum_{u' \not\in \BN} q(u, u') > 0\}
\]
as the `boundary' of~$\BN$ and $K(S',\SV') = \max_{u \in S'} \sum_{i\in \SV'} u_i$. In order to get a tight lower bound in~\eqref{propweigh} we thus need to find a subset $\BN$ such that $K(\BN,V)$ is large, $K(\partial \BN,V)$ is small, and $K(\BN,\SV)$ is small.

We will now first give an example to illustrate the use of Proposition~\ref{weighted}.
\begin{example}
\label{exmp1}
Suppose that $\BN$ is such that
$u + e_i \not\in \Omega \setminus \BN$ for all $u \in \partial \BN$.
In case $\phi_i \equiv 1$, $\psi_i \equiv 1$, $\mu_i \equiv 1$
and $\nu_i \equiv \nu \geq 1$, we then have
$Q(\BN) \leq N \pi(\partial \BN)$, and thus using~\eqref{statdistU},
\begin{align*}
\frac{1}{\Phi(\BN)}&=\frac{\pi(\BN)}{Q(\BN)} \geq \frac{\sum\limits_{u \in \BN} \pi(u)}
{N \sum\limits_{u \in \partial \BN} \pi(u)} =
\frac{\sum\limits_{u \in \BN} \nu^{\sumiN u_i}}
{N \sum\limits_{u \in \partial \BN} \nu^{\sumiN u_i}} %\\
%&
\geq \frac{1}{N} \nu^{K(\BN,V) - K(\partial \BN,V)}.
\end{align*}
We thus see that in this example we indeed need to choose $\BN$ such that $K(\BN,V) - K(\partial \BN,V)$ is maximized.

Now suppose the interference graph is a symmetric complete bipartite graph. That is, the nodes in $V_1=\{1,\dots,N/2\}$ interfere with, and only with, the nodes in $V_2=\{N/2+1,\dots,N\}$. In this case we have $K(\BN,V)\leq N/2$. Further, as $\BN$ is such that $u + e_i \not\in \Omega \setminus \BN$, we have $\BN = \Omega$ if and only if $K(\partial \BN,V)=0$. Thus, because $\BN = \Omega$ yields an irrelevant lower bound, we have $K(\partial \BN,V)\geq 1$.

Assuming that $\SV \subseteq V_1$ it is clear that $K(\BN,\SV)=0$
if $\BN$ only contains states where nodes in $V_2$ are active.
Hence in this case we should choose
$\BN = \{u \in \Omega: \sum_{i \in V_2} u_i \geq 1\}$,
the set of activity states where at least one of the nodes in $V_2$
is active, as this gives $K(\BN,V)=N/2$, $K(\partial \BN,V)=1$
and $K(\BN,\SV)=0$.
We thus see that the delay grows at least as fast as $\nu^{N/2-1}$.
\end{example}

As mixing times are typically long when transitions between dominant activity states are slow, it is likely that we can construct a lower bound for the mixing time that is similar to~\eqref{propweigh}.
The mixing time of a process represents the amount of time required for the process to come close to its equilibrium distribution,
and is formally defined as
\[
t_{{\rm mix}}(\epsilon) = \inf\{t : d(t) \leq \epsilon\},
\]
where $d(t)$ denotes the maximal distance (in total variation) between
$U(t)$ and $\pi$, i.e.,
\[
d(t) = \max_{U(0)\in\Omega} \frac{1}{2} \sum_{u \in \Omega} |\pr{U(t)=u} - \pi(u)|.
\]

As the next proposition shows, the bottleneck ratio $\Phi(\cdot)$ provides a lower
bound on the mixing time of the activity process $\{U(t)\}$.
\begin{proposition}
\label{lbmixtime}
The mixing time of $\{U(t)\}$ satisfies
\begin{equation}
\label{boundmixtime}
t_{{\rm mix}}(\epsilon) \geq \max_{\BN \subseteq \Omega} (1-2\epsilon-\pi(\BN)) \frac{1}{\Phi(\BN)}.
\end{equation}
\end{proposition}

\textbf{Proof}
Zocca {\em et al.}~\cite{ZBvL12} show that
\[
t_{{\rm mix}}(\epsilon) \geq (1-2\epsilon-r) \frac{1}{\Phi^{*}_r},
\]
with
\[
\Phi^{*}_r = \min_{\{\BN \subseteq \Omega : \pi(\BN) \leq r\}} \Phi(\BN).
\]
Therefore,
\begin{align*}
t_{{\rm mix}}(\epsilon) &\geq \max_r \max_{\{\BN \subseteq \Omega : \pi(\BN) \leq r\}} (1-2\epsilon-r) \frac{1}{\Phi(\BN)} \\
&= \max_r \max_{\{\BN \subseteq \Omega : \pi(\BN) = r\}} (1-2\epsilon-r) \frac{1}{\Phi(\BN)},
\end{align*}
and~\eqref{boundmixtime} follows.
\qed

We thus found a lower bound for the mixing time that has a similar form as the bound we found in Proposition~\ref{weighted} for the aggregate weighted queue length. Note, however, that to find a tight lower bound for the mixing time, for sufficiently small $\epsilon$, we only need $K(\BN,V)$ to be large and $K(\partial \BN,V)$ to be small. %Further note that the mixing time of networks of the type discussed in Example~\ref{exmp1} were studied in~\cite{ZBvL12} and it was shown that the lower bound in Example~\ref{exmp1} is tight.

\section{Complete partite graphs}
\label{sscompl}

In the previous section we derived generic lower bounds for the
expected aggregate weighted queue length and delays in terms of the bottleneck ratio of any subset $\BN \subseteq \Omega$, an approach that is also used to find a lower bound for the mixing time of the activity process $\{U(t)\}$.
In this section and the next we describe how to find a subset $\BN \subseteq \Omega$ with the desired properties discussed in the previous section, for a broad class of $K$-partite interference graphs. We additionally assume that each of the nodes belongs to at least one clique of size~$K$ (of which the other $K - 1$ nodes necessarily belong to $K - 1$ different components).

%The above class of graphs covers a wide range of network topologies
%with nearest-neighbor interference, e.g., linear topologies, ring
%networks with an even number of nodes, two-dimensional grid networks,
%tori (two-dimensional grid networks with a wrap-around boundary)
%with an even number of nodes in both directions, and {\em complete\/}
%$K$-partite graphs, where all nodes are connected except those that
%belong to the same component. Star networks for example have complete partite interference graphs.

We first introduce some further notation and state a few preparatory
lemmas.
Denote by $V_k \subseteq V$ the subset of nodes that belong to the
$k$-th component and $M_k = |V_k|$, $k = 1, \dots, K$.
For compactness, define
\[
\Upsilon_k = \prod\limits_{i \in V_k} (1 + \sigma_i) - 1 =
\sum_{I \subseteq V_k} \prod\limits_{i \in I} \sigma_i - 1 =
\sum_{\emptyset \neq I \subseteq V_k} \prod\limits_{i \in I} \sigma_i.
\]
In particular when $\sigma_i \equiv \hat\sigma_k$ for all $i \in V_k$,
we have $\Upsilon_k = (1 + \hat\sigma_k)^{M_k} - 1$.

Throughout we assume that $\rho_i = \hat\rho_k$ for all $i \in V_k$,
and denote $\rho = \sum_{k = 1}^{K} \hat\rho_k$,
and $\rho_{\min} = \min_{k = 1, \dots, K} \hat\rho_k$.
For convenience, we also assume $\phi_i \equiv 1$, $\psi_i \equiv 1$,
$\mu_i \equiv 1$, so that $\sigma_i = \nu_i$ for all $i = 1, \dots, N$.
Define $M = \max_{k = 1, \dots, K} M_k$ as the maximum component size.

In order to gain some useful intuition, we focus first on
{\em complete\/} $K$-partite graphs, where all nodes are connected
except those that belong to the same component.
In other words, the complement of the graph consists of $K$~fully
connected components.
Thus, transmission activity is mutually exclusive across the
various components.

In this case, the normalization constant in~\eqref{statdistU} satisfies
\[
Z = 1 + \sum_{k = 1}^{K} \sum_{\emptyset \neq I \subseteq V_k}
\prod\limits_{i \in I} \sigma_i = 1 + \sum_{k = 1}^{K} \Upsilon_k.
\]
%For any ${\mathcal V} \subseteq V$, denote by $p({\mathcal V}) = \sum_{u \in \Omega} \pi(u) \max\limits_{i \in {\mathcal V}} u_i$
%the probability that at least one of the nodes in~${\mathcal V}$ is active.
%Note that $p(V_k) = \Upsilon_k / Z$.
For any $k = 1, \dots, K$, define
$\BN_k = \{u \in \Omega: \sum_{i \in V_k} u_i \geq 1\}$ as the set of
activity states where at least one of the nodes in~$V_k$ is active.
We will use these sets to find a lower bound for the delay and mixing time.
As discussed in Example~\ref{exmp1}, these sets are likely to provide
a tight lower bound.

\begin{lemma}
\label{Qcomplete}
For any activation rate vector $(\nu_1, \dots, \nu_N)$ such that the
system is stable, for any $k = 1, \dots, K$,
\begin{align}
Q(\BN_k) &= Q(\Omega \setminus \BN_k) < M_k (1 - \sum_{l \neq k} \hat\rho_l)\left(\frac{1-\rho}{\hat\rho_k}\right)^{M_k - 1}, \label{Qcompleteeq} \\
\hat\rho_k &< \pi(\BN_k) < 1 - \sum_{l \neq k} \hat\rho_l  \label{picompleteSk}, \\
\sum_{l \neq k} \hat\rho_l &< \pi(\Omega \setminus \BN_k) < 1 - \hat\rho_k \label{picompleteSigmaminSk}.
\end{align}
\end{lemma}

\textbf{Proof}
Using~\eqref{statdistU} we obtain
\begin{align*}
\theta_i &= Z^{- 1} \sum\limits_{u \in \Omega, u_i = 1} \prod\limits_{j = 1}^{N} \sigma_j^{u_j}%\\
%&
\geq
Z^{- 1} \sigma_i \prod\limits_{l \in V_k \setminus \{i\}} (1 + \sigma_l) =
Z^{- 1} \frac{\sigma_i}{1 + \sigma_i} (\Upsilon_k + 1).
\end{align*}
Also, from~\eqref{statdistU} we know
\[
\theta_i=\sigma_i \pr{U_j = 0 \mbox{ for all } j \in \calN_i^+}.
\]
Furthermore,
\[
\pr{U_j = 0 \mbox{ for all } j \in \calN_i^+} \leq \pr{U_j = 0 \mbox{ for all } j \in \calC},
\]
and hence we get
\[
\theta_i \leq \sigma_i [1 - \sum_{j \in \calC} \theta_j],
\]
from~\eqref{actinclique}, which may be rewritten as
\[
\theta_i \leq \frac{\sigma_i}{1 + \sigma_i}
[1 - \sum_{j \in \calC \setminus \{i\}} \theta_j],
\]
so that
\[
\Upsilon_k \leq Z [1 - \sum_{j \in \calC \setminus \{i\}} \theta_j] - 1,
\]
and thus, using the fact that $\rho_i < \theta_i$ for all $i \in V$ is a necessary condition for stability,
\begin{equation}
\Upsilon_k <
Z [1 - \sum_{j \in \calC \setminus \{i\}} \rho_j] - 1.
\label{ineq6}
\end{equation}
Next, note that $Q(\Omega \setminus \BN_k) = \pi(0)  \sum_{i \in V_k} \sigma_i$, and similarly,
\[
Q(\BN_k) = \pi(0)  \sum_{i \in V_k} \sigma_i = \frac{1}{Z} \sum_{i \in V_k} \sigma_i.
\]
Using this we get,
\begin{align*}
Q(\BN_k) &\leq \frac{M_k}{Z} \max_{i \in V_k} \sigma_i =
M_k \max_{i \in V_k} \frac{\frac{\sigma_i}{1 + \sigma_i} (\Upsilon_k + 1)}
{\frac{1}{1 + \sigma_i} (\Upsilon_k + 1)} %\\
%&
 =
\frac{M_k}{Z} \max_{i \in V_k} \frac{\frac{\sigma_i}{1 + \sigma_i} (\Upsilon_k + 1)}
{\prod\limits_{l \in V_k \setminus \{i\}} (1 + \sigma_l)} \\
&<
\frac{M_k(\Upsilon_k + 1)}
{Z(1+\min_{i \in V_k} \sigma_i)^{M_k - 1}}.
\end{align*}
Invoking Lemma~\ref{lbactfac} and~\eqref{ineq6} gives~\eqref{Qcompleteeq}.

Also, because $\hat\rho_k < \pi(\BN_k) = \Upsilon_k / Z$ is needed for stability,~\eqref{ineq6} gives,
\[
\pi(\BN_k) < 1 - \sum_{l \neq k} \hat\rho_l - \frac{1}{Z},
\]
which proves~\eqref{picompleteSk}. Noting that $\pi(\BN_k)+\pi(\Omega \setminus \BN_k)=1$ gives~\eqref{picompleteSigmaminSk}.
\qed

Using Lemma~\ref{Qcomplete} we can find a lower bound for the expected aggregate weighted queue length at some subset of nodes in $\SV \subseteq V_k$.
\begin{theorem}
\label{complete}
For any activation rate vector $(\nu_1, \dots, \nu_N)$ such that the system is stable and for any $w \in {\mathbb R}_+^N$, $\SV \subseteq V_k$,
\[
\sum_{i \in \SV} w_i \expect{\TP_i} >
\frac{1}{2 M} (\rho_{\min})^{M + 1} \sum_{i \in \SV} w_i \lambda_i
\left(\frac{1}{1 - \rho}\right)^{M - 1}.
\]
For the symmetric scenario $M_k \equiv M$
and $\hat\rho_k \equiv \rho / K$ for all $k = 1, \dots, K$,
\[
\expect{\TP_i} >
\frac{(K - 1)^2 \rho^{M + 2}}{2 M K^{M + 1} (K - (K - 1) \rho)}
\left(\frac{1}{1 - \rho}\right)^{M - 1}.
\]
\end{theorem}

\textbf{Proof}
The proof relies on applying Proposition~\ref{weighted},
taking $\BN$ to be (i) $\Omega \setminus S_k$ and (ii) $S_l$, $l \neq k$.
In either case, $u_i = 0$ for all $i \in \SV$, $u \in \BN$,
so that $Y(w, \SV, \BN) = 0$, i.e.,
\[
D(w, \SV, \BN) = \sum_{i \in \SV} \lambda_i w_i.
\]
First consider case~(i). In this case we obtain the lower bound
\[
\sum_{i \in \SV} w_i \expect{\TP_i} >
\frac{\left(\sum_{l \neq k} \hat\rho_l\right)^2}{2 M_k}
\sum_{i \in \SV} w_i \lambda_i
\left(\frac{\hat\rho_k}{1 - \rho}\right)^{M_k - 1}
\]
from Proposition~\ref{weighted} and Lemma~\ref{Qcomplete}.
Taking $\SV = V_k$ yields the second statement of the lemma for
a symmetric scenario.

In order to complete the proof of the first part of the lemma,
we now turn to case~(ii).
Using Proposition~\ref{weighted} and Lemma~\ref{Qcomplete},
we arrive at the lower bound
\[
\sum_{i \in \SV} w_i \expect{\TP_i} >
\frac{\hat\rho_l^2}{2 M_l} \sum_{i \in \SV} w_i \lambda_i
\left(\frac{\hat\rho_l}{1 - \rho}\right)^{M_l - 1}.
\]
Combining the above two lower bounds yields the first part of the lemma.
\qed

Theorem~\ref{complete} states that in a complete $K$-partite graph the
expected queue length grows at least as fast as $1/(1-\rho)^{M-1}$,
with $M$ the size of the largest component.
Based on the observations in Section~\ref{prel}, this may be
heuristically explained as follows.
In order for the system to be stable, each node must at least have
an activation rate of the order $1 / (1 - \rho)$, see Lemma~\ref{lbactfac}.
In turn, the transition times between the various activity states as
governed by the maximum-size component occur on a time scale of the
order $\nu^{M-1}$, when each node has a fixed activation rate~$\nu$.

For $M = 1$ (full interference graph), the lower bound established in
Theorem~\ref{complete} is loose, reflecting that it is not the slow
transitions between the various components that cause the delays to be
long in that case, but the sheer load.
For $M = 2$, the lower bound could also have been obtained by treating
cliques as single-resource systems and is in fact similar to~\eqref{LBFC}.
For $M \geq 3$, the lower bound is particularly relevant, and reflects
that the slow transitions between the various components cause the
delays to be exponentially larger than can be explained from sheer
load considerations alone.

Lemma~\ref{Qcomplete} also provides a corresponding lower bound for
the mixing time of the activity process $\{U(t)\}$ as established in
the next theorem.

\begin{theorem}
\label{mixtimecomplete}
For any activation rate vector $(\nu_1, \dots, \nu_N)$ such that the system is stable,
\[
t_{{\rm mix}}(\epsilon) > ((K-1)\rho_{\min} - 2\epsilon) \frac{(\rho_{\min})^M}{M}  \left(\frac{1}{1 - \rho}\right)^{M - 1}.
\]
\end{theorem}

\textbf{Proof}
Applying Proposition~\ref{lbmixtime} for $\BN = \BN_k$ and using Lemma~\ref{Qcomplete} gives
\begin{align*}
t_{{\rm mix}}(\epsilon) &> (1-2\epsilon-(1 - \sum_{l \neq k} \hat\rho_l)) \frac{\hat\rho_k}{M_k (1 - \sum_{l \neq k} \hat\rho_l)}\left(\frac{\hat\rho_k}{1-\rho}\right)^{M_k - 1}\\
&> ((K-1)\rho_{\min}-2\epsilon) \frac{\rho_{\min}}{M_k}\left(\frac{\rho_{\min}}{1-\rho}\right)^{M_k - 1},
\end{align*}
for $k = 1, \dots, K$, and the result follows.
\qed

Note that the bound derived in Theorem~\ref{mixtimecomplete} is not
necessarily the tightest bound we can find using the results of
Lemma~\ref{Qcomplete}.
In fact, the bound is irrelevant if $\epsilon > (K-1)\rho_{\min}/2$.
However, if $\rho_{\min}>0$ and $\epsilon$ is small enough,
we can conclude that the mixing time grows at least like
$1/(1 - \rho)^{M - 1}$ as $\rho$ increases to~1.

For equal activation rates, i.e., for the activation rate vector
$(\nu, \dots, \nu)$, it is shown in \cite{ZBvL12} that
$t_{{\rm mix}}(\epsilon) \sim \nu^{M^{*}-1}$ as $\nu \to \infty$,
with $M^{*}$ the size of the second largest component, so that the
heavy-traffic behavior is governed by $M^{*}$ instead of~$M$.
Note however that this activation rate vector does not provide
a stable system, unless $\rho_{\min}=0$ or $M=M^{*}$, as follows from
Lemma~\ref{Qcomplete} and the fact that $\pi(\BN_k)\rightarrow 0$
as $\nu \rightarrow \infty$ for all $k$ such that $M_k < M$.
Hence the activity process mixes slower in heavy traffic if the system
is stable as compared to a system with equal activation rates.

\section{Extensions}
\label{exte}

In this section we turn attention to the broader class of (not
necessarily complete) $K$-partite graphs.
Thus, transmission activity is no longer mutually exclusive across
the various components.
However, we make the next assumption implying that joint activity
across various components is relatively inefficient.
Denote by $v^{(k)} = 1_{V_k}$ the incidence vector of~$V_k$, i.e.,
$v_i^{(k)} = 1$ if $i \in V_k$ and $v_i^{(k)} = 0$ otherwise,
and define $\Omega^* = \{v^{(1)}, \dots, v^{(K)}\}$.

\begin{assumption}
\label{assu1}
For any $u \in \Omega$,
\[
H(u) = \sum_{k = 1}^{K} \sum_{i \in V_k} \frac{u_i}{M_k} \leq 1,
\]
with strict inequality for any $u \not\in \Omega^*$.
\end{assumption}
Based on the above assumption, we define
\[
\zeta = 1 - \max_{u \in \Omega\setminus\Omega^*} H(u) > 0.
\]
An illustrative example is provided by a $2 B \times 2 B$ grid with
nodes labeled as $\{(i, j)\}$, $i, j = 1, \dots, 2 B$,
and nearest-neighbor interference.
The two components are $V_1 = \{(i, j): (i + j) \mbox{ mod } 2 = 1\}$
and $V_2 = \{(i, j): (i + j) \mbox{ mod } 2 = 0\}$, with
$M_1 = M_2 = 2 B^2$.
In order for $m \geq 1$ nodes in~$V_1$ to be active, at least
$m + 1$ or $m + 3$ nodes in~$V_2$ must be inactive (depending on
whether or not we assume a wrap-around boundary).
Thus $\sum_{i = 1}^{2 B} \sum_{j = 1}^{2 B} u_{(i, j)} \leq 2 B^2 - 1$
(or $2 B^2 - 3$) for all $u \in \Omega \setminus \Omega^*$,
and $\zeta = \frac{1}{2 B^2}$ (or $\frac{3}{2 B^2}$).

The next lemma shows that in order for the system to be stable,
joint activity across the various components can only occur
a negligible fraction of the time at high load.
%The proof of this lemma is deferred to Appendix~\ref{proofnegligible}.
\begin{lemma}
\label{negligible}
In order for the system to be stable, it should hold that
\[
\sum_{u \in \Omega \setminus \Omega^*} \pi(u) < \frac{1 - \rho}{\zeta},
\]
and
\[
\pi(v^{(k)}) > \hat\rho_k - \frac{1 - \rho}{\zeta}.
\]
\end{lemma}

\textbf{Proof}
In order for the system to be stable, we must have $\rho_i < \theta_i$
for all $i = 1, \dots, N$.
Thus,
\begin{align*}
\rho &= \sum_{k = 1}^{K} \hat\rho_k= \sum_{k = 1}^{K} \frac{1}{M_k} \sum_{i \in V_k} \rho_i \\
&<\sum_{k = 1}^{K} \frac{1}{M_k} \sum_{i \in V_k} \sum_{u \in \Omega} \pi(u) u_i%\\
%&
=
\sum_{u \in \Omega} \pi(u) \sum_{k = 1}^{K} \frac{1}{M_k} \sum_{i \in V_k} u_i \\
&=\sum_{u \in \Omega^*} \pi(u) \sum_{k = 1}^{K} \frac{1}{M_k} \sum_{i \in V_k} u_i +
\sum_{u \in \Omega \setminus \Omega^*} \pi(u) \sum_{k = 1}^{K} \frac{1}{M_k}
\sum_{i \in V_k} u_i \\
&\leq\sum_{u \in \Omega^*} \pi(u) +
(1 - \zeta) \sum_{u \in \Omega \setminus \Omega^*} \pi(u)\\
&=
1 - \zeta \sum_{u \in \Omega \setminus \Omega^*} \pi(u),
\end{align*}
where the last inequality follows from Assumption~\ref{assu1}.
The first part of the lemma follows.

Also, for any $i \in V_k$,
\[
\hat\rho_k = \rho_i < \sum_{u \in \Omega} \pi(u) u_i \leq
\pi(v^{(k)}) + \sum_{u \in \Omega \setminus \Omega^*} \pi(u),
\]
which combined with the first statement yields the second part of the lemma.
\qed

In the next lemma %, whose proof can be found in Appendix~\ref{proofcloseratio},
we show that the fraction of time the activity process $\{U(t)\}$ spends in any component $V_k$ relative to the traffic intensity of the nodes in that component, is almost equal for all components if $\rho$ is large enough.
\begin{lemma}
\label{closeratio}
Assume the system is stable and $\rho \geq \rho_{\gamma} = 1 - \gamma \zeta \rho_{\min}^2$, $\gamma>0$.
Then
\[
\min_{k = 1, \dots, K} \frac{1}{\hat\rho_k} \prod\limits_{i \in V_k} \sigma_i \geq
(1 - 3 \gamma)
\max_{k = 1, \dots, K} \frac{1}{\hat\rho_k} \prod\limits_{i \in V_k} \sigma_i.
\]
\end{lemma}

\textbf{Proof}
For compactness, we denote $\Pi_k = \prod_{i \in V_k} \sigma_i$
and $R_k = \Pi_k / \hat\rho_k$,
and define $k_{\min} = \arg\min_{k = 1, \dots, K} R_k$
and $k_{\max} = \arg\max_{k = 1, \dots, K} R_k$.

%We need to show that $R_{k_{\max}} \leq (1 + 3 \gamma) R_{k_{\min}}$.

Lemma~\ref{negligible} implies
\begin{align*}
\hat\rho_{k_{\min}} - \gamma \rho_{\min}^2 &\leq
\hat\rho_{k_{\min}} - (1 - \rho) / \zeta \leq \pi(v^{(k_{\min})}) =
Z^{- 1} \Pi_{k_{\min}} \\
&\leq
\frac{\Pi_{k_{\min}}}{\sum_{k = 1}^{K} \Pi_k} %\\
%&
=
\frac{\Pi_{k_{\min}}}{\Pi_{k_{\min}} + \Pi_{k_{\max}} +
\sum_{k \neq k_{\min}, k_{\max}} \Pi_k} \\
&\leq
\frac{\Pi_{k_{\min}}}{\Pi_{k_{\min}} + \Pi_{k_{\max}} +
\sum_{k \neq k_{\min}, k_{\max}}
\frac{\hat\rho_{k}}{\hat\rho_{k_{\min}}} \Pi_{k_{\min}}},
\end{align*}
yielding
\[
(1 - (\hat\rho_{k_{\min}} + \sum_{k \neq k_{\min}, k_{\max}} \hat\rho_k)
(1 - \frac{\gamma \rho_{\min}^2}{\hat\rho_{k_{\min}}})) \frac{\Pi_{k_{\min}}}{\Pi_{k_{\max}}} \geq
\hat\rho_{k_{\min}} - \gamma \rho_{\min}^2,
\]
or equivalently,
\[
(1 - (\rho - \hat\rho_{k_{\max}})
(1 - \gamma \rho_{\min}^2 / \hat\rho_{k_{\min}})) \Pi_{k_{\min}} \geq
(\hat\rho_{k_{\min}} - \gamma \rho_{\min}^2) \Pi_{k_{\max}}.
\]
Using $\rho \geq 1 - \gamma \zeta \rho_{\min}^2$
and $\rho_{\min}\leq \hat\rho_{k_{\min}}$, it follows that
\[
(1 - (1 - \gamma \zeta \rho_{\min}^2 - \hat\rho_{k_{\max}})
(1 - \gamma \rho_{\min})) \Pi_{k_{\min}} \geq
(1 - \gamma) \hat\rho_{k_{\min}} \Pi_{k_{\max}}.
\]
This yields
\[
(1 + 2 \gamma) \hat\rho_{k_{\max}} \Pi_{k_{\min}} \geq
(1 - \gamma) \hat\rho_{k_{\min}} \Pi_{k_{\max}},
\]
and thus,
\[
R_{k_{\min}} \geq \frac{(1 - \gamma) R_{k_{\max}}}{1 + 2 \gamma} \geq
(1 - 3 \gamma) R_{k_{\max}}.
\]
\qed

In order to state a lower bound for the expected aggregate weighted
queue length at some subset of nodes $\SV \subseteq V_k$,
we now first introduce some further notation and concepts.

A sequence of states $(u^{(0)}, u^{(1)}, \dots, u^{(l)})$, with
$u^{(k)} \in \Omega$, $k = 0, \dots, l$, is called a \textit{path}
from~$u^{(0)}$ to~$u^{(l)}$ if $(u^{(k)}, u^{(k+1)})$ are feasible
transitions, i.e., $q(u^{(k)}, u^{(k+1)}) > 0$ for all
$k = 0, \dots, l - 1$.
For a given path $p = (u^{(0)}, u^{(1)}, \dots, u^{(l)})$, denote
by $m(p) = \min_{k = 0, 1, \dots, l} H(u^{(k)})$ the minimum value
of the function $H(\cdot)$, as defined in Assumption~\ref{assu1}, along the path.
For given states $u, v \in \Omega$, denote by $P(u, v)$ the collection
of all paths from~$u$ to~$v$.
Define $M(u, v) = \max_{p \in P(u, v)} m(p)$ as the maximum of the
minimum value of the function $H(\cdot)$ along any path from
state~$u$ to state~$v$, with the convention that $M(u, u) = \infty$.

For all $\SV \subseteq V$ such that $\SV \subseteq V_k$ for some $k \in \{1,\dots,K\}$, denote by $\Delta(\SV)$ the set of states in which the expected drift of the aggregate weighted queue length in $\SV$ is non-positive, i.e.,
\[
\Delta(\SV)= \{u \in \Omega: \sum_{i\in \SV} w_i \lambda_i \leq \sum_{i\in \SV} w_i \mu_i u_i\}.
\]
Further define $\delta(\SV)$ as the minimal expected drift of the aggregate weighted queue length in $\SV$ if the system does not reside in of one of the states in $\Delta(\SV)$, i.e.,
\[
\delta(\SV) = \sum_{i\in \SV} w_i \lambda_i - \max_{u \in \Omega \setminus \Delta(\SV)} \sum_{i\in \SV} w_i \mu_i u_i.
\]
Note that $\delta(\SV)>0$ by construction.
For all $l \neq k$, define $m_l(\SV) = \max_{u \in \Delta(\SV)} M(v^{(l)}, u)$,
and
\[
\BN_l(\SV) = \{u \in \Omega: M(v^{(l)}, u) > m_l(\SV)\}
\]
as the set of states that can be reached from $v^{(l)}$ via
a path~$p$ with $m(p) > m_l(\SV)$.
%By definition $\BN_l \cap \Delta(\SV) = \emptyset$, and thus
%$D(w, \SV, \BN_l) \geq \delta(\SV)$, $l \neq k$.
Also, define $m_k(\SV) = \max_{l \neq k} m_l(\SV)$, and
\[
\BN_k(\SV) = \{v \in \Omega: \max_{u \in \Delta(\SV)} M(u, v) > m_k(\SV)\}
\]
as the set of states that can be reached from $\Delta(\SV)$ via a path~$p$
with $m(p) > m_k(\SV)$.
%Note that $\Delta(\SV) \subseteq \BN_k$, so that
%$D(w, \SV, \Omega \setminus \BN_k) \geq \delta(\SV)$,
%and $\BN_k \cap \BN_l = \emptyset$, i.e., $\BN_l \subseteq \Omega \setminus \BN_k$,
%$l \neq k$.

Finally, define $H_l(\SV) = \max_{u \in \partial S_l(\SV)} H(u)$, $H^*(\SV) =\min_{l = 1, \dots, K} H_l(\SV)$ and $H^*_{\min} = \min_{\SV \subseteq V: \exists k: \SV \subseteq V_k} H^*(\SV)$.

In the remainder of this subsection we will assume that the activation rates of nodes in the same component are equal, i.e.~$\sigma_i=\hat\sigma_k$ if $i\in V_k$ for all $k=1,\dots,K$. Denote $\sigma^* = \min_{k = 1, \dots, K} \hat\sigma_k^{M_k}$ and $k^{*}=\argmin_{k = 1, \dots, K} \hat\sigma_k^{M_k}$.
\begin{remark}
It is not clear when there exists an activation rate vector $(\nu_1, \dots, \nu_N)$ with $\nu_i=\nu_j$ if $i,j\in V_k$ that stabilizes the system. For symmetric topologies, e.g.~ring networks with an even number of nodes or tori with an even number of nodes in both directions, it seems plausible that such an activation rate vector can stabilize the system for any $\rho<1$. For asymmetric typologies, e.g.~linear topologies and two-dimensional grid networks, this is not clear.
\end{remark}
In the next lemma we derive an upper bound for the fraction of the time the system spends in the boundary of $S_l(\SV)$ for any $l=1,\dots,K$.
%The proof of this lemma is deferred to Appendix~\ref{proofupperbound}.
\begin{lemma}
\label{upperbound}
Assume the system is stable and $\rho \geq \rho_{\gamma} = 1 - \gamma \zeta \rho_{\min}^2$, $\gamma>0$. Then
\[
\max_{u \in \partial S_l(\SV)}
\prod\limits_{j = 1}^{N} \sigma_j^{u_j} \leq \left(\frac{\sigma^*}{(1 - 3 \gamma) \rho_{\min}}\right)^{H_l(\SV)}.
\]
\end{lemma}

\textbf{Proof}
Since $\sigma_i = \hat\sigma_k$ for all $k = 1, \dots, K$, we obtain
\begin{align*}
\max_{u \in \partial S_l(\SV)} \prod\limits_{j = 1}^{N} \sigma_j^{u_j} &=
\max_{u \in \partial S_l(\SV)}
\prod\limits_{k = 1}^{K} \prod\limits_{i \in V_k} \sigma_i^{u_i} %\\
%&
=
\max_{u \in \partial S_l(\SV)}
\prod\limits_{k = 1}^{K} \hat\sigma_k^{\sum\limits_{i \in V_k} u_i} \\
 &=
\max_{u \in \partial S_l(\SV)} \prod\limits_{k = 1}^{K}
(\hat\sigma_k^{M_k})^{\frac{1}{M_k} \sum\limits_{i \in V_k} u_i}.
\end{align*}
Lemma~\ref{closeratio} gives
\[
\hat\sigma_k^{M_k} \leq \frac{\hat\rho_k}{1 - 3 \gamma} \min_{l=1,\dots,K} \frac{1}{\hat\rho_l} \hat\sigma_l^{M_l} \leq \frac{\sigma^*}{(1 - 3 \gamma) \rho_{\min}},
\]
and thus,
\begin{align*}
\max_{u \in \partial S_l(\SV)} \prod\limits_{j = 1}^{N} \sigma_j^{u_j}&\leq
\max_{u \in \partial S_l(\SV)} \prod\limits_{k = 1}^{K}
\left(\frac{\sigma^*}{(1 - 3 \gamma) \rho_{\min}}\right)^{\frac{1}{M_k} \sum\limits_{i \in V_k} u_i}\\
&\leq
\max_{u \in \partial S_l(\SV)}
\left(\frac{\sigma^*}{(1 - 3 \gamma) \rho_{\min}}\right)^{\sum_{k = 1}^{K} \frac{1}{M_k} \sum\limits_{i \in V_k} u_i}\\
&=
\left(\frac{\sigma^*}{(1 - 3 \gamma) \rho_{\min}}\right)^{\max_{u \in \partial S_l(\SV)} H(u)}\\
&=
\left(\frac{\sigma^*}{(1 - 3 \gamma) \rho_{\min}}\right)^{H_l(\SV)}.
\end{align*}
\qed

We are now in the position to derive bounds for $Q(\BN_l(\SV))$, $\pi(\BN_l(\SV))$ and $\pi(\Omega \setminus \BN_l(\SV))$ that are qualitatively similar to the bounds in Lemma~\ref{Qcomplete}.
\begin{lemma}
\label{Qexten}
Assume $\rho \geq \rho_\gamma = 1 - \gamma \zeta \rho_{\min}^2$, $\gamma > 0$.
For any activation rate vector $(\nu_1, \dots, \nu_N)$ such that the system is stable and with $\nu_i=\nu_j$ if $i,j\in V_k$ for some $k$, for any $l = 1, \dots, K$,
\begin{align}
Q(\BN_l(\SV)) &= Q(\Omega \setminus \BN_l(\SV)) \nonumber \\
&< \frac{2^N}{(1 - 3 \gamma) \rho_{\min}}
\left(\frac{\rho_{k^*}}{1 - \rho}\right)^{M (H_l(\SV) - 1)}, \label{Qexteneq} \\
(1 - \gamma) \rho_{\min} &< \pi(\BN_l(\SV)) < 1 - (1 - \gamma) \rho_{\min} \label{piextenSk}, \\
(1 - \gamma) \rho_{\min} &< \pi(\Omega \setminus \BN_l(\SV)) < 1 - (1 - \gamma) \rho_{\min} \label{piextenSigmaminSk}.
\end{align}
\end{lemma}

\textbf{Proof}
First note that
\[
Q(\BN_l(\SV)) = \sum_{u \in \partial S_l(\SV)} \pi(u)\]
and
\[
Q(\Omega \setminus \BN_l(\SV)) = \sum_{u \in \partial S_l(\SV)} \pi(u).
\]
Further,
\begin{align*}
Q(\BN_l(\SV)) &= \sum_{u \in \partial S_l(\SV)} \pi(u) =
Z^{- 1} \sum_{u \in \partial S_l(\SV)} \prod\limits_{j = 1}^{N}
\sigma_j^{u_j} \\
&\leq
Z^{- 1} | \partial S_l(\SV) | \max_{u \in \partial S_l(\SV)}
\prod\limits_{j = 1}^{N} \sigma_j^{u_j}.
\end{align*}
Noting that $Z \geq \sigma^*$ and $| \partial S_l(\SV) | \leq 2^N$ yields, using Lemma~\ref{upperbound},
\[
Q(\BN_l(\SV)) \leq
\frac{2^N \left(\frac{\sigma^*}{(1 - 3 \gamma) \rho_{\min}}\right)^{H_l(\SV)}}{\sigma^*} \leq
\frac{2^N (\sigma^*)^{H_l(\SV)-1}}{(1 - 3 \gamma) \rho_{\min}},
\]
and~\eqref{Qexteneq} follows from Lemma~\ref{lbactfac}.

Further, using Lemma~\ref{negligible},
\[
\pi(\BN_l(\SV)) = \pi(S_l(\SV)) \geq \pi(v^{(l)}) >
\hat\rho_l - \frac{1 - \rho}{\zeta} \geq (1 - \gamma) \rho_{\min}.
\]
Now note that by definition $\BN_l \cap \Delta(\SV) = \emptyset$ for $l\neq k$ and $\Delta(\SV) \subseteq \BN_k$ if $\SV \subseteq V_k$. Hence, for $l\neq k$,
\[
\pi(\BN_l(\SV)) \leq 1 - \pi(v^{(k)}) < 1- (1 - \gamma) \rho_{\min},
\]
and
\[
\pi(\BN_k(\SV)) \leq 1 - \sum_{l \neq k} \pi(v^{(l)}) < 1- (K-1) (1 - \gamma) \rho_{\min},
%&\leq 1- (1 - \gamma) \rho_{\min},
\]
which gives~\eqref{piextenSk}. Noting that $\pi(\BN_l(\SV))+\pi(\Omega \setminus \BN_l(\SV))=1$ gives~\eqref{piextenSigmaminSk}.
\qed

Using a similar approach as in Section~\ref{sscompl}, the bounds in
Lemma~\ref{Qexten} can be utilized to establish a lower bound for the
expected aggregate weighted queue length in some subset of nodes
and for the mixing time of the activity process.
\begin{theorem}
\label{extensions}
Assume $\rho \geq \rho_\gamma = 1 - \gamma \zeta \rho_{\min}^2$, $\gamma > 0$.
For any activation rate vector $(\nu_1, \dots, \nu_N)$, with $\nu_i=\nu_j$ if $i,j\in V_k$ for some $k$, such that the system is stable and for any $w \in {\mathbb R}_+^N$, $\SV \subseteq V_k$,
\[
\sum_{i \in \SV} w_i \expect{\TP_i} > \frac{\delta(\SV) (1 - 4 \gamma) \rho_{\min}^{M + 3}}{2^{N + 1}}
\left(\frac{1}{1 - \rho}\right)^{M (1 - H^*(\SV))},
\]
\end{theorem}

\textbf{Proof}
The proof of this theorem proceeds along similar lines as the proof of Theorem~\ref{complete} and relies on applying Proposition~\ref{weighted}, taking $\BN$ to be (i) $\Omega \setminus S_k(\SV)$ and (ii) $\BN = S_l(\SV)$, $l \neq k$. First note that by definition $\BN_l(\SV) \cap \Delta(\SV) = \emptyset$, and thus $D(w, \SV, \BN_l(\SV)) \geq \delta(\SV)$, $l \neq k$. Also note that $\Delta(\SV) \subseteq \BN_k(\SV)$, so that $D(w, \SV, \Omega \setminus \BN_k(\SV)) \geq \delta(\SV)$.

Further, using Lemma~\ref{Qexten} we obtain the lower bound
\[
\sum_{i \in \SV} w_i \expect{\TP_i} >
\frac{\delta(\SV) (1 - 4 \gamma) \rho_{\min}^{M + 3}}{2^{N + 1}}
\left(\frac{1}{1 - \rho}\right)^{M (1 - H_l(\SV))},
\]
for $l=1,\dots,K$, and the result follows.
\qed

Theorem~\ref{extensions} states that in a general $K$-partite
interference graph the expected queue length grows at least as fast as
$1/(1-\rho)^{M(1-H^*)}$, where the coefficient $H^*$ depends on the
specific topology and is in general hard to calculate.
We however know that $\frac{1}{M} \leq H^* \leq 1$ and for some
specific topologies we can explicitly determine $H^*$.

The next theorem provides a corresponding lower bound for the mixing
time of the activity process $\{U(t)\}$.

\begin{theorem}
\label{mixtimeexten}
Assume $\rho \geq \rho_\gamma = 1 - \gamma \zeta \rho_{\min}^2$, $\gamma > 0$.
For any activation rate vector $(\nu_1, \dots, \nu_N)$ such that the system is stable and with $\nu_i=\nu_j$ if $i,j\in V_k$ for some $k$,
\[
t_{{\rm mix}}(\epsilon) > ((1 - \gamma) \rho_{\min}-2\epsilon) \frac{(1 - 4\gamma) \rho_{\min}^{M+2}}{2^N}
\left(\frac{1}{1 - \rho}\right)^{M (1-H^*_{\min})}.
\]
\end{theorem}

\textbf{Proof}
%From Lemma~\ref{lbmixtime} we know
%\[
%t_{{\rm mix}}(\epsilon) \geq (1-2\epsilon-\pi(\BN)) \frac{1}{\Phi(\BN)},
%\]
%for any $\BN \subseteq \Omega$.
Take $\SV \subseteq V$ such that there exists a $k$ such that $\SV \subseteq V_k$. Using Lemma~\ref{Qcomplete} we then find for any $l\in \{1,\dots,K\}$,
\begin{align*}
\frac{1}{\Phi(\BN_l(\SV))} &> \frac{(1 - \gamma) \rho_{\min}(1 - 3 \gamma) \rho_{\min}}{2^N}
\left(\frac{\rho_{k^*}}{1 - \rho}\right)^{M (1-H_l(\SV))} \\
&\geq \frac{(1 - 4\gamma) \rho_{\min}^{M+2}}{2^N}
\left(\frac{1}{1 - \rho}\right)^{M (1-H_l(\SV))}.
\end{align*}
Hence, using Proposition~\ref{lbmixtime},
\[
t_{{\rm mix}}(\epsilon) > ((1 - \gamma) \rho_{\min}-2\epsilon) \frac{(1 - 4\gamma) \rho_{\min}^{M+2}}{2^N}
\left(\frac{1}{1 - \rho}\right)^{M (1-H_l(\SV))},
\]
and the result follows by optimizing over~$l$ and~$\SV$.
\qed

The value of the coefficient~$H^*(\SV)$ depends strongly on the
specific properties of the interference graph~$G$.
For a complete partite graph, the sets $S_l(\SV)$ coincide with those
in the previous subsection, and we have
$\partial S_l(\SV) = \bigcup_{i \in V_l} \{e_i\}$, so that
$H_l(\SV) = 1 / M_l$, and $H^*(\SV) = 1 / M$, recovering the result of
Theorem~\ref{complete}.
On the other hand, when the graph consists of $N / K$ fully connected
components, we have $H_l(\SV) \equiv 1$, and the result trivializes.
An interesting intermediate situation is the $2 B \times 2 B$ grid
mentioned earlier with $M = M_1 = M_2 = 2 B^2$, for which we
conjecture that $H^*(\SV) = H_1 = H_2 = 1 - 1 / B$ or $1 - 1 / (2 B)$ if $B\geq 2$,
depending on whether or not we assume a wrap-around boundary,
suggesting that the mean queue lengths would grow as
$1 / (1 - \rho)^B$ or $1 / (1 - \rho)^{2 B}$.

\section{Simulation experiments}
\label{simu}

In this section we will illustrate the theoretical results for the growth
behavior of the aggregate queue length through simulation experiments.
For cross comparison, we consider a system that can be represented
by a symmetric complete bipartite ($K=2$) interference graph
with components of size $M=5$.
Because of space considerations, we do not report simulation results
for other cases, but we observed qualitatively similar behavior in
a broad range of scenarios.

To estimate the expected aggregate queue length for a given value
of~$\rho$, we set $t=10^6$ and calculate the average total number
of packets in the time intervals $[0,t]$ and $[t+1,2t]$,
starting from an initially empty system.
We take the average of the two values to be our estimate if the values
are less than $5\%$ apart.
Otherwise we set $t=2t$ and repeat the procedure.

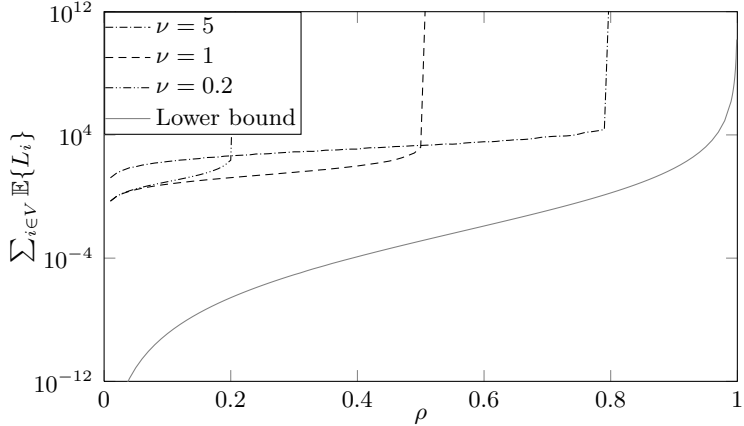
\begin{figure}
  \centering
  \begin{tikzpicture}%
\begin{semilogyaxis}[
font=\footnotesize,
width=10cm, height=6.5cm,
    xmin=0, xmax=1,
    ymin=1e-12, ymax=1e12,
    scaled ticks=false,
    every axis x label/.style=
        {at={(ticklabel cs:0.5)},anchor=center},
	xlabel={$\rho$},
    every axis y label/.style=
        {at={(ticklabel cs:0.5)},rotate=90,anchor=center},
    ylabel={$\sum_{i\in V} \expect{L_i}$},
    legend cell align=left,
    legend style={at={(0,1)}, anchor=north west, inner xsep=0pt, inner ysep=0pt}]

\legend{
	$\nu=5$\\
    $\nu=1$\\
	$\nu=0.2$\\%
    Lower bound\\%
}
\addplot[color=black,densely dashdotted] plot coordinates {
	(0.010000,	16.391823)
	(0.020000,	33.742053)
	(0.030000,	51.360854)
	(0.040000,	73.236328)
	(0.050000,	86.467494)
	(0.060000,	109.191678)
	(0.070000,	127.161519)
	(0.080000,	149.288701)
	(0.090000,	164.274036)
	(0.100000,	186.824688)
	(0.110000,	206.700724)
	(0.120000,	231.100981)
	(0.130000,	246.165386)
	(0.140000,	270.434315)
	(0.150000,	289.407838)
	(0.160000,	337.009268)
	(0.170000,	348.893233)
	(0.180000,	366.412433)
	(0.190000,	415.022801)
	(0.200000,	436.464353)
	(0.210000,	465.226843)
	(0.220000,	486.062587)
	(0.230000,	507.694667)
	(0.240000,	578.168464)
	(0.250000,	576.320169)
	(0.260000,	601.305692)
	(0.270000,	656.735990)
	(0.280000,	695.226404)
	(0.290000,	740.867529)
	(0.300000,	798.312286)
	(0.310000,	797.228035)
	(0.320000,	858.274540)
	(0.330000,	852.662172)
	(0.340000,	917.887822)
	(0.350000,	1016.320274)
	(0.360000,	1027.265542)
	(0.370000,	1067.399460)
	(0.380000,	1157.705236)
	(0.390000,	1194.996927)
	(0.400000,	1193.288704)
	(0.410000,	1354.593275)
	(0.420000,	1462.766573)
	(0.430000,	1460.348295)
	(0.440000,	1551.398820)
	(0.450000,	1630.902242)
	(0.460000,	1758.353927)
	(0.470000,	1791.112812)
	(0.480000,	1896.008323)
	(0.490000,	2008.003286)
	(0.500000,	2061.982571)
	(0.510000,	2160.268015)
	(0.520000,	2221.961684)
	(0.530000,	2417.996652)
	(0.540000,	2572.029955)
	(0.550000,	2496.357901)
	(0.560000,	2642.164201)
	(0.570000,	2982.828091)
	(0.580000,	3198.490350)
	(0.590000,	3367.990648)
	(0.600000,	3601.739941)
	(0.610000,	3561.015780)
	(0.620000,	3807.164442)
	(0.630000,	4273.745076)
	(0.640000,	4456.461802)
	(0.650000,	4911.805015)
	(0.660000,	4958.790986)
	(0.670000,	5327.413667)
	(0.680000,	5928.599044)
	(0.690000,	6938.783356)
	(0.700000,	7546.021219)
	(0.710000,	8067.928774)
	(0.720000,	8315.135752)
	(0.730000,	8586.489288)
	(0.740000,	9373.879008)
	(0.750000,	11333.977173)
	(0.760000,	14608.071691)
	(0.770000,	16132.293015)
	(0.780000,	18621.134702)
	(0.790000,	22546.872928)
	(0.800000,	9999999999999999)
};
\addplot[color=black,densely dashed] plot coordinates {
	(0.010000,	0.506356)
	(0.020000,	1.041311)
	(0.030000,	1.597451)
	(0.040000,	2.182585)
	(0.050000,	2.785648)
	(0.060000,	3.458098)
	(0.070000,	4.094785)
	(0.080000,	4.844718)
	(0.090000,	5.570195)
	(0.100000,	6.301414)
	(0.110000,	7.096828)
	(0.120000,	7.971194)
	(0.130000,	8.831907)
	(0.140000,	9.832363)
	(0.150000,	10.788079)
	(0.160000,	11.894510)
	(0.170000,	13.012186)
	(0.180000,	14.141083)
	(0.190000,	15.456933)
	(0.200000,	16.749748)
	(0.210000,	18.150844)
	(0.220000,	19.784341)
	(0.230000,	21.321659)
	(0.240000,	23.104205)
	(0.250000,	25.166766)
	(0.260000,	26.867373)
	(0.270000,	29.061572)
	(0.280000,	31.790505)
	(0.290000,	34.642919)
	(0.300000,	37.156419)
	(0.310000,	40.869607)
	(0.320000,	44.033474)
	(0.330000,	47.703501)
	(0.340000,	52.808531)
	(0.350000,	56.811575)
	(0.360000,	63.008478)
	(0.370000,	69.463507)
	(0.380000,	77.269357)
	(0.390000,	85.891283)
	(0.400000,	95.845411)
	(0.410000,	109.821938)
	(0.420000,	126.227371)
	(0.430000,	144.154749)
	(0.440000,	168.428654)
	(0.450000,	198.279472)
	(0.460000,	246.505423)
	(0.470000,	324.800502)
	(0.480000,	459.557835)
	(0.490000,	671.579179)
	(0.500000,	1592.211283)
	(0.510000,	9999999999999999)
};
\addplot[color=black,densely dashdotdotted] plot coordinates {
	(0.010000,	0.499255)
	(0.020000,	1.053440)
	(0.030000,	1.669063)
	(0.040000,	2.354930)
	(0.050000,	3.145174)
	(0.060000,	4.006452)
	(0.070000,	5.039967)
	(0.080000,	6.201765)
	(0.090000,	7.519031)
	(0.100000,	9.119675)
	(0.110000,	11.074334)
	(0.120000,	13.365228)
	(0.130000,	16.383090)
	(0.140000,	20.333400)
	(0.150000,	25.302499)
	(0.160000,	32.758981)
	(0.170000,	44.139169)
	(0.180000,	63.335066)
	(0.190000,	102.784282)
	(0.200000,	229.362188)
	(0.210000,	9999999999999999)
};
\addplot[color=gray] plot coordinates {
	(0.010000,	8.173840732453024e-17)
	(0.020000,	1.09511695427671e-14)
	(0.030000,	1.95936635265618e-13)
	(0.040000,	1.53779683169564e-12)
	(0.050000,	7.68565744663227e-12)
	(0.060000,	2.88779017426771e-11)
	(0.070000,	8.91286242284017e-11)
	(0.080000,	2.38230518997097e-10)
	(0.090000,	5.70582773305574e-10)
	(0.100000,	1.25341932792658e-09)
	(0.110000,	2.56771602413468e-09)
	(0.120000,	4.965925621287932e-09)
	(0.130000,	9.151765324962467e-09)
	(0.140000,	1.618860873166650e-08)
	(0.150000,	2.764466987910796e-08)
	(0.160000,	4.578518120580909e-08)
	(0.170000,	7.382424539027013e-08)
	(0.180000,	1.162520885990954e-07)
	(0.190000,	1.792571892056694e-07)
	(0.200000,	2.712673611111112e-07)
	(0.210000,	4.036395311196764e-07)
	(0.220000,	5.915349081954377e-07)
	(0.230000,	8.550247989414693e-07)
	(0.240000,	1.220482849690730e-06)
	(0.250000,	1.722332451499118e-06)
	(0.260000,	2.405234001504809e-06)
	(0.270000,	3.326816383049861e-06)
	(0.280000,	4.561081292556795e-06)
	(0.290000,	6.202639007534257e-06)
	(0.300000,	8.371971237474582e-06)
	(0.310000,	0.000011)
	(0.320000,	0.000015)
	(0.330000,	0.000020)
	(0.340000,	0.000026)
	(0.350000,	0.000034)
	(0.360000,	0.000045)
	(0.370000,	0.000058)
	(0.380000,	0.000075)
	(0.390000,	0.000096)
	(0.400000,	0.000123)
	(0.410000,	0.000158)
	(0.420000,	0.000201)
	(0.430000,	0.000256)
	(0.440000,	0.000325)
	(0.450000,	0.000412)
	(0.460000,	0.000520)
	(0.470000,	0.000656)
	(0.480000,	0.000825)
	(0.490000,	0.001037)
	(0.500000,	0.001302)
	(0.510000,	0.001632)
	(0.520000,	0.002045)
	(0.530000,	0.002559)
	(0.540000,	0.003200)
	(0.550000,	0.004001)
	(0.560000,	0.005000)
	(0.570000,	0.006248)
	(0.580000,	0.007808)
	(0.590000,	0.009760)
	(0.600000,	0.012204)
	(0.610000,	0.015271)
	(0.620000,	0.019123)
	(0.630000,	0.023970)
	(0.640000,	0.030084)
	(0.650000,	0.037810)
	(0.660000,	0.047600)
	(0.670000,	0.060039)
	(0.680000,	0.075894)
	(0.690000,	0.096171)
	(0.700000,	0.122202)
	(0.710000,	0.155757)
	(0.720000,	0.199207)
	(0.730000,	0.255753)
	(0.740000,	0.329745)
	(0.750000,	0.427148)
	(0.760000,	0.556222)
	(0.770000,	0.728515)
	(0.780000,	0.960355)
	(0.790000,	1.275108)
	(0.800000,	1.706667)
	(0.810000,	2.304909)
	(0.820000,	3.144452)
	(0.830000,	4.338955)
	(0.840000,	6.065088)
	(0.850000,	8.603787)
	(0.860000,	12.413441)
	(0.870000,	18.264296)
	(0.880000,	27.495127)
	(0.890000,	42.526178)
	(0.900000,	67.939901)
	(0.910000,	112.905016)
	(0.920000,	197.038510)
	(0.930000,	365.952508)
	(0.940000,	737.571512)
	(0.950000,	1662.707848)
	(0.960000,	4410.068677)
	(0.970000,	15132.125075)
	(0.980000,	83115.572653)
	(0.990000,	1441932.778321)
	(0.991000,	2215513.513851)
	(0.992000,	3577513.250751)
	(0.993000,	6152379.761668)
	(0.994000,	11490045.131375)
	(0.995000,	24017926.537549)
	(0.996000,	59110103.101906)
	(0.997000,	188321620.368428)
	(0.998000,	961050282.793129)
	(0.999000,	15500452126.544950)
};
\end{semilogyaxis}
\end{tikzpicture}%
  \caption{Average total number of packets for several fixed activation rates.}
  \label{fixedact}
\end{figure}

Figure~\ref{fixedact} shows the average total number of packets in the
system for various fixed activation rates.
Note that we used a log-lin scale.
We see that the simulated curves lie well above the lower bound
of Theorem~\ref{complete} for all chosen values of~$\nu$.
Note that the system is not stable for all values of~$\rho$, e.g.~for
$\nu=1$ the system is unstable if $\rho\geq2\theta_i=\frac{32}{63}$,
explaining the jumps in the simulation result.
Further note that the expected time between activation of nodes
in the two components is smaller for small values of~$\nu$.
This explains why small values of~$\nu$ tend to perform better in case
$\rho$ is small, i.e., for large values of~$\nu$ the nodes in one
component will often be transmitting dummy packets while the nodes
in the other component do have packets waiting to be transmitted.

%\begin{figure}
%  \centering
%  \includegraphics[width=3.3in,keepaspectratio]{flog2.png}
%  \caption{Average total number of packets for a symmetric complete bipartite graph with $M=4$ for $f(\tp)=\log(\tp+1)$ and $g(\tp)=1$.}
%  \label{logact}
%\end{figure}

\begin{figure}
  \centering
  \begin{tikzpicture}%
\begin{semilogyaxis}[
font=\footnotesize,
width=10cm, height=6.5cm,
    xmin=0, xmax=1,
    ymin=1e-9, ymax=1e9,
    scaled ticks=false,
    every axis x label/.style=
        {at={(ticklabel cs:0.5)},anchor=center},
	xlabel={$\rho$},
    every axis y label/.style=
        {at={(ticklabel cs:0.5)},rotate=90,anchor=center},
    ylabel={$\sum_{i\in V} \expect{L_i}$},
    legend cell align=left,
    legend style={at={(0,1)}, anchor=north west, inner xsep=0pt, inner ysep=0pt}]

\legend{%
	Simulation for $f(\tp)=\tp$ and $g(\tp)=1$\\%
	Lower bound for $f(\tp)=\tp$ and $g(\tp)=1$\\%
	Lower bound for any fixed activation rate vector\\%
}%
\addplot[color=black] plot coordinates {
	(0.010000,	1.028777e-001)
	(0.020000,	2.136243e-001)
	(0.030000,	3.281267e-001)
	(0.040000,	4.524877e-001)
	(0.050000,	5.819702e-001)
	(0.060000,	7.227999e-001)
	(0.070000,	8.673682e-001)
	(0.080000,	1.024208e+000)
	(0.090000,	1.187618e+000)
	(0.100000,	1.363207e+000)
	(0.110000,	1.545683e+000)
	(0.120000,	1.743501e+000)
	(0.130000,	1.941690e+000)
	(0.140000,	2.160604e+000)
	(0.150000,	2.389742e+000)
	(0.160000,	2.619594e+000)
	(0.170000,	2.882570e+000)
	(0.180000,	3.140622e+000)
	(0.190000,	3.413341e+000)
	(0.200000,	3.698857e+000)
	(0.210000,	4.008939e+000)
	(0.220000,	4.317178e+000)
	(0.230000,	4.651865e+000)
	(0.240000,	5.005982e+000)
	(0.250000,	5.351715e+000)
	(0.260000,	5.735373e+000)
	(0.270000,	6.125146e+000)
	(0.280000,	6.525399e+000)
	(0.290000,	6.953147e+000)
	(0.300000,	7.376445e+000)
	(0.310000,	7.837075e+000)
	(0.320000,	8.334995e+000)
	(0.330000,	8.813102e+000)
	(0.340000,	9.323404e+000)
	(0.350000,	9.887503e+000)
	(0.360000,	1.046617e+001)
	(0.370000,	1.102509e+001)
	(0.380000,	1.164943e+001)
	(0.390000,	1.229470e+001)
	(0.400000,	1.294658e+001)
	(0.410000,	1.363048e+001)
	(0.420000,	1.437655e+001)
	(0.430000,	1.514599e+001)
	(0.440000,	1.599431e+001)
	(0.450000,	1.676815e+001)
	(0.460000,	1.759973e+001)
	(0.470000,	1.854233e+001)
	(0.480000,	1.952454e+001)
	(0.490000,	2.052276e+001)
	(0.500000,	2.160428e+001)
	(0.510000,	2.269612e+001)
	(0.520000,	2.392201e+001)
	(0.530000,	2.513113e+001)
	(0.540000,	2.649599e+001)
	(0.550000,	2.788489e+001)
	(0.560000,	2.943153e+001)
	(0.570000,	3.098257e+001)
	(0.580000,	3.255726e+001)
	(0.590000,	3.439409e+001)
	(0.600000,	3.616926e+001)
	(0.610000,	3.836345e+001)
	(0.620000,	4.048800e+001)
	(0.630000,	4.292289e+001)
	(0.640000,	4.559362e+001)
	(0.650000,	4.811368e+001)
	(0.660000,	5.131329e+001)
	(0.670000,	5.416337e+001)
	(0.680000,	5.745157e+001)
	(0.690000,	6.095068e+001)
	(0.700000,	6.539430e+001)
	(0.710000,	6.996886e+001)
	(0.720000,	7.480480e+001)
	(0.730000,	7.989763e+001)
	(0.740000,	8.609066e+001)
	(0.750000,	9.335726e+001)
	(0.760000,	1.004352e+002)
	(0.770000,	1.094029e+002)
	(0.780000,	1.176299e+002)
	(0.790000,	1.277756e+002)
	(0.800000,	1.416663e+002)
	(0.810000,	1.540598e+002)
	(0.820000,	1.686824e+002)
	(0.830000,	1.873669e+002)
	(0.840000,	2.079960e+002)
	(0.850000,	2.351482e+002)
	(0.860000,	2.665731e+002)
	(0.870000,	3.010080e+002)
	(0.880000,	3.398465e+002)
	(0.890000,	3.999295e+002)
	(0.900000,	4.719782e+002)
	(0.910000,	5.782957e+002)
	(0.920000,	6.949013e+002)
	(0.930000,	8.663959e+002)
	(0.940000,	1.140158e+003)
	(0.950000,	1.549854e+003)
	(0.960000,	2.221232e+003)
	(0.970000,	3.681198e+003)
	(0.980000,	7.737798e+003)
	(0.990000,	2.477437e+004)
	(0.991000,	2.923935e+004)
	(0.992000,	3.707290e+004)
	(0.993000,	4.488493e+004)
	(0.994000,	6.144011e+004)
	(0.995000,	8.341479e+004)
	(0.996000,	1.342828e+005)
	(0.997000,	2.328830e+005)
};
\addplot[color=black,densely dashed] plot coordinates {
	(0.010000,	1.010101e-001)
	(0.020000,	2.040816e-001)
	(0.030000,	3.092784e-001)
	(0.040000,	4.166667e-001)
	(0.050000,	5.263158e-001)
	(0.060000,	6.382979e-001)
	(0.070000,	7.526882e-001)
	(0.080000,	8.695652e-001)
	(0.090000,	9.890110e-001)
	(0.100000,	1.111111e+000)
	(0.110000,	1.235955e+000)
	(0.120000,	1.363636e+000)
	(0.130000,	1.494253e+000)
	(0.140000,	1.627907e+000)
	(0.150000,	1.764706e+000)
	(0.160000,	1.904762e+000)
	(0.170000,	2.048193e+000)
	(0.180000,	2.195122e+000)
	(0.190000,	2.345679e+000)
	(0.200000,	2.500000e+000)
	(0.210000,	2.658228e+000)
	(0.220000,	2.820513e+000)
	(0.230000,	2.987013e+000)
	(0.240000,	3.157895e+000)
	(0.250000,	3.333333e+000)
	(0.260000,	3.513514e+000)
	(0.270000,	3.698630e+000)
	(0.280000,	3.888889e+000)
	(0.290000,	4.084507e+000)
	(0.300000,	4.285714e+000)
	(0.310000,	4.492754e+000)
	(0.320000,	4.705882e+000)
	(0.330000,	4.925373e+000)
	(0.340000,	5.151515e+000)
	(0.350000,	5.384615e+000)
	(0.360000,	5.625000e+000)
	(0.370000,	5.873016e+000)
	(0.380000,	6.129032e+000)
	(0.390000,	6.393443e+000)
	(0.400000,	6.666667e+000)
	(0.410000,	6.949153e+000)
	(0.420000,	7.241379e+000)
	(0.430000,	7.543860e+000)
	(0.440000,	7.857143e+000)
	(0.450000,	8.181818e+000)
	(0.460000,	8.518519e+000)
	(0.470000,	8.867925e+000)
	(0.480000,	9.230769e+000)
	(0.490000,	9.607843e+000)
	(0.500000,	1.000000e+001)
	(0.510000,	1.040816e+001)
	(0.520000,	1.083333e+001)
	(0.530000,	1.127660e+001)
	(0.540000,	1.173913e+001)
	(0.550000,	1.222222e+001)
	(0.560000,	1.272727e+001)
	(0.570000,	1.325581e+001)
	(0.580000,	1.380952e+001)
	(0.590000,	1.439024e+001)
	(0.600000,	1.500000e+001)
	(0.610000,	1.564103e+001)
	(0.620000,	1.631579e+001)
	(0.630000,	1.702703e+001)
	(0.640000,	1.777778e+001)
	(0.650000,	1.857143e+001)
	(0.660000,	1.941176e+001)
	(0.670000,	2.030303e+001)
	(0.680000,	2.125000e+001)
	(0.690000,	2.225806e+001)
	(0.700000,	2.333333e+001)
	(0.710000,	2.448276e+001)
	(0.720000,	2.571429e+001)
	(0.730000,	2.703704e+001)
	(0.740000,	2.846154e+001)
	(0.750000,	3.000000e+001)
	(0.760000,	3.166667e+001)
	(0.770000,	3.347826e+001)
	(0.780000,	3.545455e+001)
	(0.790000,	3.761905e+001)
	(0.800000,	4.000000e+001)
	(0.810000,	4.263158e+001)
	(0.820000,	4.555556e+001)
	(0.830000,	4.882353e+001)
	(0.840000,	5.250000e+001)
	(0.850000,	5.666667e+001)
	(0.860000,	6.142857e+001)
	(0.870000,	6.692308e+001)
	(0.880000,	7.333333e+001)
	(0.890000,	8.090909e+001)
	(0.900000,	9.000000e+001)
	(0.910000,	1.011111e+002)
	(0.920000,	1.150000e+002)
	(0.930000,	1.328571e+002)
	(0.940000,	1.566667e+002)
	(0.950000,	1.900000e+002)
	(0.960000,	2.400000e+002)
	(0.970000,	3.233333e+002)
	(0.980000,	4.900000e+002)
	(0.990000,	9.900000e+002)
	(0.991000,	1.101111e+003)
	(0.992000,	1.240000e+003)
	(0.993000,	1.418571e+003)
	(0.994000,	1.656667e+003)
	(0.995000,	1.990000e+003)
	(0.996000,	2.490000e+003)
	(0.997000,	3.323333e+003)
};
\addplot[color=gray,densely dashdotdotted] plot coordinates {
	(0.010000,	8.173841e-018)
	(0.020000,	1.095117e-015)
	(0.030000,	1.959366e-014)
	(0.040000,	1.537797e-013)
	(0.050000,	7.685657e-013)
	(0.060000,	2.887790e-012)
	(0.070000,	8.912862e-012)
	(0.080000,	2.382305e-011)
	(0.090000,	5.705828e-011)
	(0.100000,	1.253419e-010)
	(0.110000,	2.567716e-010)
	(0.120000,	4.965926e-010)
	(0.130000,	9.151765e-010)
	(0.140000,	1.618861e-009)
	(0.150000,	2.764467e-009)
	(0.160000,	4.578518e-009)
	(0.170000,	7.382425e-009)
	(0.180000,	1.162521e-008)
	(0.190000,	1.792572e-008)
	(0.200000,	2.712674e-008)
	(0.210000,	4.036395e-008)
	(0.220000,	5.915349e-008)
	(0.230000,	8.550248e-008)
	(0.240000,	1.220483e-007)
	(0.250000,	1.722332e-007)
	(0.260000,	2.405234e-007)
	(0.270000,	3.326816e-007)
	(0.280000,	4.561081e-007)
	(0.290000,	6.202639e-007)
	(0.300000,	8.371971e-007)
	(0.310000,	1.122196e-006)
	(0.320000,	1.494600e-006)
	(0.330000,	1.978800e-006)
	(0.340000,	2.605485e-006)
	(0.350000,	3.413178e-006)
	(0.360000,	4.450141e-006)
	(0.370000,	5.776733e-006)
	(0.380000,	7.468329e-006)
	(0.390000,	9.618937e-006)
	(0.400000,	1.234568e-005)
	(0.410000,	1.579435e-005)
	(0.420000,	2.014633e-005)
	(0.430000,	2.562714e-005)
	(0.440000,	3.251712e-005)
	(0.450000,	4.116472e-005)
	(0.460000,	5.200308e-005)
	(0.470000,	6.557075e-005)
	(0.480000,	8.253758e-005)
	(0.490000,	1.037373e-004)
	(0.500000,	1.302083e-004)
	(0.510000,	1.632453e-004)
	(0.520000,	2.044640e-004)
	(0.530000,	2.558835e-004)
	(0.540000,	3.200311e-004)
	(0.550000,	4.000744e-004)
	(0.560000,	4.999920e-004)
	(0.570000,	6.247901e-004)
	(0.580000,	7.807832e-004)
	(0.590000,	9.759545e-004)
	(0.600000,	1.220424e-003)
	(0.610000,	1.527058e-003)
	(0.620000,	1.912266e-003)
	(0.630000,	2.397046e-003)
	(0.640000,	3.008368e-003)
	(0.650000,	3.781008e-003)
	(0.660000,	4.760002e-003)
	(0.670000,	6.003939e-003)
	(0.680000,	7.589420e-003)
	(0.690000,	9.617132e-003)
	(0.700000,	1.222019e-002)
	(0.710000,	1.557567e-002)
	(0.720000,	1.992074e-002)
	(0.730000,	2.557533e-002)
	(0.740000,	3.297449e-002)
	(0.750000,	4.271484e-002)
	(0.760000,	5.562225e-002)
	(0.770000,	7.285152e-002)
	(0.780000,	9.603548e-002)
	(0.790000,	1.275108e-001)
	(0.800000,	1.706667e-001)
	(0.810000,	2.304909e-001)
	(0.820000,	3.144452e-001)
	(0.830000,	4.338955e-001)
	(0.840000,	6.065088e-001)
	(0.850000,	8.603787e-001)
	(0.860000,	1.241344e+000)
	(0.870000,	1.826430e+000)
	(0.880000,	2.749513e+000)
	(0.890000,	4.252618e+000)
	(0.900000,	6.793990e+000)
	(0.910000,	1.129050e+001)
	(0.920000,	1.970385e+001)
	(0.930000,	3.659525e+001)
	(0.940000,	7.375715e+001)
	(0.950000,	1.662708e+002)
	(0.960000,	4.410069e+002)
	(0.970000,	1.513213e+003)
	(0.980000,	8.311557e+003)
	(0.990000,	1.441933e+005)
	(0.991000,	2.215514e+005)
	(0.992000,	3.577513e+005)
	(0.993000,	6.152380e+005)
	(0.994000,	1.149005e+006)
	(0.995000,	2.401793e+006)
	(0.996000,	5.911010e+006)
	(0.997000,	1.883216e+007)
};
\end{semilogyaxis}
\end{tikzpicture}%
  \caption{Average total number of packets for $f(\tp)=\tp$ and $g(\tp)=1$.}
  \label{xact}
\end{figure}
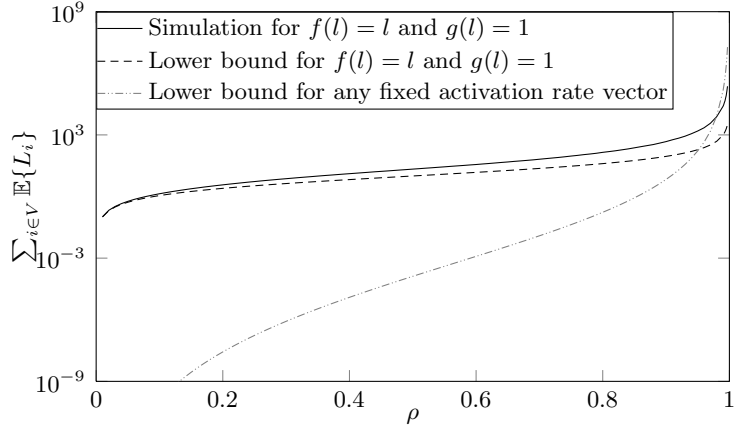

Figure~\ref{xact} shows the average total number of packets in the
system for $f(\tp)=\tp$ and $g(\tp)=1$.
We see that the lower bound of Theorem~\ref{backlogb} is remarkably
close to the simulation result for small values of~$\rho$.
For larger values of~$\rho$ the bound and simulation result are
farther apart.
One explanation for this lies in the approximation made in~\eqref{boundact}.
For small values of~$\rho$ this approximation is relatively good
while for large values of~$\rho$ this approximation is off by a factor
of about~2 in this case.
While this does not explain the total discrepancy in this case,
it does explain all discrepancies in case the rate of increase of the
activation function is slow, e.g.~$f(\tp)=\log(\tp+1)$.

Finally note that the simulation result lies, for large values of~$\rho$,
below the lower bound for fixed activation rates established
in Theorem~\ref{complete}.
This suggests that the activation function $f_i(\tp)\equiv f(\tp)=\tp$
performs better in heavy traffic than $f_i(\tp)=\nu_i$ for any choice
of the activation rate vector $(\nu_1,\dots,\nu_N)$.

\section{Conclusions}
\label{concl}

We have established lower bounds for the expected queue lengths
and delays in wireless random-access networks.
%The bounds indicate that the delay for fixed activation rates can
%dramatically grow with the load in certain topologies.
%We further showed that the delay for queue-based activation
%and de-activation strategies exceeds this lower bound when the degree
%of aggressiveness grows slowly as function of the queue length.
%An interesting issue for further research is to determine when better
%delay performance can be achieved when the aggressiveness grows faster
%as function of the queue length.
Both for queue-based strategies and fixed activation rates, the
derivation of the bounds starts from the observation that stability of
the system requires the activity factors to be big at high load.
The specific subsequent arguments considerably differ however in both
cases.
Queue-based strategies for which maximum stability has been established,
involve slow, logarithmic, activation functions, which require huge
queue lengths at every node for the activity factors to be big enough,
and cause the exponential delay scaling.
In contrast, the delays for fixed activation rates are shown to result
from excessive mixing times due to a bottleneck in the network topology
together with the big activity factors required for stability.
We also observe that the network topology plays a major role in case
of fixed activation rates, while it only appears to matter somewhat
implicitly in case of queue-based strategies as will be further
discussed below.
%It is worth emphasizing that the activation rule for which maximum
%stability is guaranteed, however, {\em does\/} depend on the network
%topology.
%Also, the lower bounds rely on the presence of (critically-loaded) cliques,
%and hence the network topology does implicitly play a role in the case
%of queue-based strategies as well.
%We further see that, in heavy traffic, the lower bound for general partite graphs grows at most as fast as the lower bound we found for complete $K$-partite interference graphs.
%This suggests that a complete $K$-partite interference graph is the worst topology in terms of delay performance if one uses a fixed-rate strategy and therefore it is interesting to analyze this topology.

For complete partite interference graphs, a comparison of both cases
reveals that the expected delay for queue-based strategies grows
faster than the lower bound $1 / (1 - \rho)^{M - 1}$ for fixed
activation and de-activation rates when $h(\tp)$ increases slower than
$\tp^{1 / (M - 1)}$, with $M$ denoting the maximum component size.
This is for example the case if $f(\tp) = r(\tp) / (1 + r(\tp))$
and $g(\tp) = 1 / (1 + r(\tp))$, with $r(\tp) = \log(\tp + 1)$.
Conversely, when $h(\tp)$ increases faster than $\tp^{1 / (M - 1)}$,
the lower bound for fixed activation and de-activation rates could
potentially be beaten by sufficiently aggressive queue-based strategies.
Simulation experiments demonstrate that the actual expected delays
indeed exhibit the cross-over suggested by the lower bounds.

A challenging issue for further research is to examine whether more
aggressive queue-based strategies can improve the delay performance
in more general topologies as well.
As noted earlier, maximum-stability guarantees in arbitrary topologies
have only been established so far for nominal activation functions
that grow logarithmically with the queue lengths \cite{RSS09a,SS12,SST11}.
Inspection of the proof arguments indicates that maximum stability
will remain guaranteed as long as the fluid limits of the queue length
process exhibit fast mixing behavior.
This in turn means that the activity process for such queue-based
strategies in fact behaves as if the activation rates are essentially
fixed.
Thus, in arbitrary topologies it is questionable whether queue-based
strategies have the capability to outperform fixed-rate strategies.

In some specific topologies, however, maximum stability is maintained
for highly aggressive queue-based strategies for which the fluid
limits of the queue length process may exhibit slow mixing behavior
\cite{FPR10,GBW12}.
The complete partite interference graphs considered in the present
paper are crucial examples of such topologies.
In these scenarios there seems to be scope for more aggressive
queue-based strategies to reduce the delays, as confirmed by the lower
bounds and simulation results that we presented.

In conclusion, the question in what kind of scenarios more aggressive
queue-based strategies can improve the delay performance appears to be
inextricably linked to the question under what conditions such
strategies provide maximum-stability guarantees.
In both these questions, the mixing properties of the activity process
seem to play a central role, and it would be interesting to explore
this three-way connection further.

\bibliographystyle{plain}
\bibliography{bibtex}

%Non-BibTeX users please use
%\begin{thebibliography}{}
%
%and use \bibitem to create references. Consult the Instructions
%for authors for reference list style.
%
%\bibitem{RefJ}
%Format for Journal Reference
%Author, Article title, Journal, Volume, page numbers (year)
%Format for books
%\bibitem{RefB}
%Author, Book title, page numbers. Publisher, place (year)
%etc
%\end{thebibliography}

%\appendix
%
%Appendix A
%
%\section{Proofs}
%\label{appendix}
%
%\subsection{Proof Lemma~\ref{negligible}}
%\label{proofnegligible}
%
%\subsection{Proof Lemma~\ref{closeratio}}
%\label{proofcloseratio}
%
%\subsection{Proof Lemma~\ref{upperbound}}
%\label{proofupperbound}

\end{document}